\pdfoutput=1
\documentclass[a4paper,11pt]{article}
\usepackage{jheppub,undertilde}

\newcommand{\be}{\begin{equation}}
\newcommand{\ee}{\end{equation}}
\newcommand\sss{\scriptscriptstyle}
\newcommand{\mW}{m_{\sss W}}
\newcommand{\sW}{s_{\sss W}}
\newcommand{\cW}{c_{\sss W}}
\newcommand{\hc}{+\,\mathrm{h.c.}}
\def\bsp#1\esp{\begin{split}#1\end{split}}
\def\bpm{\begin{pmatrix}}
\def\epm{\end{pmatrix}}
\def\ie{{\it i.e.}}
\def\eg{{\it e.g.}}

\preprint{CERN-PH-TH/2013-248}
\title{Phenomenology of the Higgs Effective Lagrangian via {\sc FeynRules}}

\author[a]{Adam Alloul,}
\author[b,c]{Benjamin Fuks}
\author[d]{and Ver\'onica Sanz}
\affiliation[a]{Groupe de Recherche de Physique des Hautes \'Energies (GRPHE), Universit\'e de Haute-Alsace, IUT Colmar, 34 rue du Grillenbreit BP 50568, 68008 Colmar Cedex, France}
\affiliation[b]{Theory Division, Physics Department, CERN, CH-1211 Geneva 23,
  Switzerland}
\affiliation[c]{Institut Pluridisciplinaire Hubert Curien/D\'epartement
  Recherches Subatomiques,\\ Universit\'e de Strasbourg/CNRS-IN2P3,
  23 rue du Loess, F-67037 Strasbourg, France}
\affiliation[d]{Department of Physics and Astronomy, University of Sussex, Brighton BN1 9QH, UK.}

\abstract{
The Higgs discovery and the lack of any other hint for new physics favor a description of
non-standard Higgs physics in terms of an effective field theory. We present
an implementation of a general Higgs effective Lagrangian containing operators up to dimension six
in the framework of {\sc FeynRules} and provide details on the translation between the mass and
interaction bases, in particular for three- and four-point interaction vertices involving Higgs and gauge bosons.
We illustrate the strengths of this implementation by using the UFO interface of {\sc FeynRules} capable to generate
model files that can be understood by the {\sc MadGraph}~5 event generator and that have
the specificity to contain all interaction vertices, without any restriction on the number of
external legs or on the complexity of the Lorentz structures. We then investigate several new physics effects
in total  rates and differential distributions for different Higgs production modes, including
gluon fusion, associated production with a gauge boson and di-Higgs production. We finally
study contact interactions of gauge and Higgs bosons to fermions.
}
\emailAdd{adam.alloul@iphc.cnrs.fr}
\emailAdd{benjamin.fuks@cern.ch}
\emailAdd{v.sanz@sussex.ac.uk}


\begin{document}
\maketitle
\flushbottom

\section{Introduction}

The discovery of a new state featuring the characteristics of a Higgs boson~\cite{Aad:2012tfa,Chatrchyan:2012ufa}
has led to an intense effort to determine whether it consists of \textit{the} Standard Model (SM) Higgs
boson~\cite{Higgs:1964pj,Higgs:1966ev}. On the one hand, its quantum numbers and properties have been
theoretically studied in the context of the Standard Model and many of
its extensions~\cite{Dell'Aquila:1985ve,Dell'Aquila:1985vc,Dell'Aquila:1985vb,Dell'Aquila:1988rx,%
Dell'Aquila:1988fe,Choi:2002jk,Odagiri:2002nd,Buszello:2002uu,Djouadi:2005gi,Buszello:2006hf,%
Bredenstein:2006rh,BhupalDev:2007is,Godbole:2007cn,Hagiwara:2009wt,Gao:2010qx,DeRujula:2010ys,Buckley:2010jv,Englert:2010ud,%
DeSanctis:2011yc,Barger:2011cb,Kumar:2011yta,Ellis:2012wg,Coleppa:2012eh,Bolognesi:2012mm,Boughezal:2012tz,%
Stolarski:2012ps,Ellis:2012xd,Alves:2012fb,Cea:2012ud,Choi:2012yg,Avery:2012um,Geng:2012hy,Ellis:2012jv,%
Freitas:2012kw,Andersen:2012kn,Ellis:2012mj,Frank:2012wh,Englert:2012xt,Bernaciak:2012nh,Djouadi:2013yb,%
Modak:2013sb,Artoisenet:2013puc}, while on the other hand,
both Tevatron and LHC experiments have determined that the resonance is likely a scalar particle with
properties close to those expected from a Standard Model Higgs boson~\cite{Aaltonen:2013kxa,ATLAS:2013sla,%
Chatrchyan:2012jja}. In particular, dedicated studies
have tried to unravel new physics by investigating possible deviations of the couplings of the new
state to the Standard Model particles~\cite{Carmi:2012yp,Azatov:2012bz,Espinosa:2012ir,%
Giardino:2012ww,Li:2012ku,Rauch:2012wa,Ellis:2012rx,Azatov:2012rd,Klute:2012pu,%
Espinosa:2012vu,Carmi:2012zd,Dolan:2012rv,Chang:2012tb,Chang:2012gn,%
Low:2012rj,Giardino:2012dp,Montull:2012ik,Espinosa:2012im,Carmi:2012in,Banerjee:2012xc,%
Bonnet:2012nm,Plehn:2012iz,Espinosa:2012in,Djouadi:2012rh,Chang:2012bq,Belanger:2012gc,%
Djouadi:2013qya,Belanger:2013xza,Altarelli:2013lla,Banerjee:2013apa}.

Alas, no direct evidence of physics beyond the Standard Model has been
observed in the light of current data~\cite{atlaspub,cmspub}, up to some anomalous
multilepton events recorded by the CMS detector~\cite{CMS-PAS-SUS-13-002}
which could find a possible explanation from gauge-mediated
supersymmetry-breaking models~\cite{D'Hondt:2013ula}.
With this state of affairs, a practical way for investigating new physics
lies in a description based on an effective field theory valid up to a scale $\Lambda$
lying around the TeV scale. In this context, the dynamics of the elementary particles
is described through higher-dimensional operators
featuring the Standard Model fields and only constrained by the SM symmetry group.
This approach allows to characterize the properties
of the newly observed Higgs boson, and also has
the advantages to be renormalizable order by order in the $E/\Lambda$
expansion, $E$ being a typical energy scale for the processes of interest, as well
as to exhibit a small number of free parameters as the gauge symmetries and the hierarchy
among the operators of different canonical dimensions strongly constrain the set of
operators relevant for a given purpose. Consequently, the path of effective
field theories for Higgs physics has been widely explored~\cite{Burges:1983zg,%
Leung:1984ni,Buchmuller:1985jz,DeRujula:1991se,%
Hagiwara:1992eh,Hagiwara:1993ck,Hagiwara:1993qt,GonzalezGarcia:1999fq,Eboli:1999pt,Giudice:2007fh,%
Grzadkowski:2010es,Bonnet:2011yx,Biswal:2012mp,Contino:2013kra,Gainer:2013rxa,Chen:2013waa},
although one of its major
drawbacks consists of the loss of unitarity,
and thus of predictivity, for energy scales greater than
$\Lambda$.

In this work, we focus on the dominant
dimension-six operators related to Higgs physics, \ie, those involving
at least one Higgs or gauge field, and neglect possible four-fermion
interactions allowed by the SM gauge symmetries. Whilst limits on the magnitude
of the associated Wilson coefficients can be extracted from LEP and Tevatron data,
Higgs signal strengths deduced from measurements
by both the ATLAS and CMS collaborations also now imply
constraints on a subset of the allowed operators~\cite{Corbett:2012dm,Masso:2012eq,%
Corbett:2012ja,Corbett:2013pja,Dumont:2013wma,Corbett:2013hia}.
However, as more information on the Higgs-boson properties is obtained by LHC
experiments, investigating new physics effects with
a simple fit of the Higgs signal strengths
clearly becomes too naive. More statistics indeed allows one for employing
various kinematical distributions as powerful handles to look for physics beyond
the Standard Model
in the Higgs sector. In these perspectives, it is necessary to rely on
sophisticated tools that cover the implementation
of the Higgs effective operators in Monte Carlo simulation programs, the latter
further leading to a possible recasting of the experimental analyses and
a subsequent comparison of the theoretical predictions,
in the framework of an effective field theory for Higgs
physics, with experimental data.

Within the last few years, a framework
based on the {\sc FeynRules} package~\cite{Christensen:2008py,
Christensen:2010wz,Duhr:2011se,Fuks:2012im,Alloul:2013fw,
Christensen:2013aua,Alloul:2013jea,Alloul:2013bka}
has been developed in order to facilitate
the implementation (and validation) of any beyond the Standard Model theory in
multi-purpose matrix-element generators~\cite{Christensen:2009jx}. In particular,
its virtues have
been illustrated in the
context of the {\sc CalcHep}~\cite{Pukhov:1999gg,Boos:2004kh,Pukhov:2004ca,%
Belyaev:2012qa}, {\sc FeynArts}/{\sc FormCalc}~\cite{Hahn:1998yk,Hahn:2000kx,%
Hahn:2006zy,Hahn:2009bf,Agrawal:2011tm,Nejad:2013ina},
{\sc MadGraph}~\cite{Stelzer:1994ta,%
Maltoni:2002qb,Alwall:2007st,Alwall:2008pm,Alwall:2011uj},
{\sc Sherpa}~\cite{Gleisberg:2003xi,Gleisberg:2008ta} and
{\sc Whizard}~\cite{Moretti:2001zz,Kilian:2007gr} programs.
Furthermore, it also offers
the possibility to translate any particle physics
model in terms of a {\sc Python} library
under the so-called Universal FeynRules Output (UFO)
format~\cite{Degrande:2011ua}.
Previously designed model formats for Monte Carlo programs are usually
imposing restrictions on the color and/or Lorentz structures and the number
of external legs
that are allowed in the interaction vertices so that the UFO has been
conceived to overcome such limitations. It is therefore suitable for effective
field theories that often contain vertices with uncommon Lorentz structures.

We present in this work an implementation in {\sc FeynRules}
of a Higgs Effective Field Theory that includes a set
of independent dimension-six operators assumed to encompass all possible
effects of new physics on the Higgs sector. This set of operators
can further be easily complemented by
22 four-fermion operators, irrelevant for Higgs physics (at least at leading-order and
in the context of the LHC phenomenology)
and thus omitted here,
in order to achieve a full basis of independent dimension-six operators featuring
only Standard Model fields. In this way,
high-energy physics programs such as {\sc Aloha}~\cite{deAquino:2011ub},
{\sc GoSam}~\cite{Cullen:2011ac,Cullen:2011xs},
{\sc Herwig++}~\cite{Bahr:2008pv,Arnold:2012fq},
{\sc MadAnalysis}~5~\cite{Conte:2012fm,Conte:2013mea}
and {\sc MadGraph}~5, capable to fully handle a UFO model,
can be directly employed for phenomenological investigations of new physics
in the context of the effective field theory model
presented in this work. For the sake of the examples, we demonstrate the
usefulness of such an implementation by studying various beyond the Standard Model
effects in both Higgs production rates and kinematical distributions by means of the
{\sc FeynRules} and {\sc MadGraph}~5 packages and their interface through the UFO.

Our work is organized as follows. In Section~\ref{sec:HEL},
we extensively describe the operators which are included in our implementation.
Next, we move onto studying several properties of the Higgs boson as
examples of phenomenological studies that can be performed from this implementation.
Their thorough study, possibly carried at the most sophisticated
level of collider physics simulations, is however beyond the scope of this work, as well
as a careful design of experimentally allowed values for the Wilson coefficients of the various
effective operators that we briefly comment in Section~\ref{sec:bounds}.
In Section~\ref{sec:HtoVV}, we firstly focus on Higgs-boson production by gluon fusion and on the constraints
that can be possibly extracted on the custodial symmetry or from measurements of angular
distributions in the case the Higgs boson decays
via intermediate gauge bosons.
Secondly, we consider in Section~\ref{sec:Htobb} the
associated production of a Higgs boson with one weak boson and
investigate the ratio of the total production rates at center-of-mass
energies of 8~TeV and 14 TeV as well as the invariant-mass spectrum of the Higgs and vector boson system.
Thirdly, we assess the effects of some effective operators on the production of
a Higgs-boson pair in Section~\ref{dihiggs}, focusing on boosted topologies.
Finally, we study in Section~\ref{sec:HZ} contact interactions of a single Higgs field,
a single gauge boson and a fermion pair and compare the LHC reach to limits extracted from LEP data.
Our conclusions are given in Section~\ref{sec:concl}.

\section{Effective Lagrangians for a Higgs doublet}~\label{sec:HEL}
\subsection{The most general effective Higgs Lagrangian in the gauge eigenbasis}
\label{sec:HELgauge}

The Standard Model of particle physics is a quantum field theory which
describes the elementary particles and their interactions based
on the $SU(3)_c\times SU(2)_L\times U(1)_Y$ gauge symmetry.
The particle content of the model can be classified in terms of a gauge sector, consisting
of vector fields responsible for the mediation of the interactions, a chiral
sector with the matter building blocks
and eventually a Higgs sector related to the breaking of the electroweak symmetry.
We now introduce our notations and start with the gauge vector
fields lying in the adjoint representation of the relevant gauge subgroup,
\be\label{eq:gaugecontent}
  SU(3)_c \to G^a_\mu  = ({\utilde{\bf 8}},{\utilde{\bf 1}},0) \ ,\quad 
  SU(2)_L \to W^k_\mu  = ({\utilde{\bf 1}},{\utilde{\bf 3}},0) \ ,\quad 
  U(1)_Y  \to B_\mu    = ({\utilde{\bf 1}},{\utilde{\bf 1}},0) \ ,
\ee
which we show together with their full representation under the Standard Model gauge group
and explicitly indicate the adjoint gauge indices.
The chiral content of the theory is defined by three generations of
left-handed and right-handed quark ($Q_L$,
$u_R$ and $d_R$) and lepton ($L_L$ and $e_R$) fields,
\be\bsp
  &Q_L = \bpm u_L\\d_L \epm =
      \big(\utilde{\bf 3}, \utilde{\bf 2},\frac16\big) \ , 
 \quad 
  u_R = \big(\utilde{\bf 3}, \utilde{\bf 1}, \frac23\big) \ , \quad
  d_R = \big(\utilde{\bf 3}, \utilde{\bf 1},-\frac13\big) \ ,\\
  &L_L = \bpm \nu_L \\ \ell_L \epm =
     \big(\utilde{\bf 1}, \utilde{\bf 2},-\frac12\big)\ ,
  \quad 
   e_R = \big(\utilde{\bf 1}, \utilde{\bf 1},-1\big)\ ,
\esp \ee
which we again present together with their representation under the Standard Model
gauge group. Finally, the Higgs sector contains a single $SU(2)_L$ doublet of fields,
\be
  \Phi = \bpm -i G^+ \\ \frac{1}{\sqrt{2}} \Big[ v + h + i G^0\Big] \epm = 
      \big(\utilde{\bf 1}, \utilde{\bf 2},\frac12\big) \ .
\ee
With the first equality, we show the component fields of the doublet after shifting the neutral field $h$
by its vacuum expectation value $v$. Moreover, 
we have included the Goldstone bosons $G^{+,0}$ to be eaten by the weak boson to
get their longitudinal degree of freedom.

In the effective field theory-based approach that we adopt,
the usual Standard Model Lagrangian ${\cal L}_{\rm SM}$
is supplemented by higher-dimensional operators that parametrize the possible effects of
non-observed states assumed to appear
at energies larger than an effective scale identified with the
$W$-boson mass $m_W$ or equivalently with the vacuum expectation value of the Higgs field $v$.
Restricting ourselves to operators of dimension less than or equal to six,
the most general gauge-invariant Lagrangian ${\cal L}$
is known for a long time \cite{Burges:1983zg,Leung:1984ni,Buchmuller:1985jz}
and can be expressed, in a convenient basis of independent operators ${\cal O}_i$~\cite{Giudice:2007fh,Grzadkowski:2010es},
as
\be\label{eq:effL}
  {\cal L} = {\cal L}_{\rm SM}  + \sum_i \bar c_i {\cal O}_i =
    {\cal L}_{\rm SM} + {\cal L}_{\rm SILH} + {\cal L}_{CP}
   + {\cal L}_{F_1} + {\cal L}_{F_2}  +  {\cal L}_{G}\ ,
\ee
assuming baryon and lepton number conservation. Moreover, we
adopt the decomposition of Ref.~\cite{Contino:2013kra} and
normalize the Wilson coefficients $\bar c_i$ as such. 
Other choices for the operator basis yield
different physics interpretations and care must be taken when
comparing various works on the topic. According to the purposes,
other bases might be more adequate.

The first piece of this Lagrangian,  ${\cal L}_{\rm SILH}$, corresponds to the
a popular set of $CP$-conserving
operators involving the Higgs doublet $\Phi$. Their
(conventional) normalization
is inspired by scenarios where the Higgs field is part of a strongly
interacting sector,
\be \label{eq:silh}\bsp
  {\cal L}_{\rm SILH} =&\
    \frac{\bar c_{\sss H}}{2 v^2} \partial^\mu\big[\Phi^\dag \Phi\big] \partial_\mu \big[ \Phi^\dagger \Phi \big]
  + \frac{\bar c_{\sss T}}{2 v^2} \big[ \Phi^\dag {\overleftrightarrow{D}}^\mu \Phi \big] \big[ \Phi^\dag {\overleftrightarrow{D}}_\mu \Phi \big] 
  - \frac{\bar c_{\sss 6} \lambda}{v^2} \big[\Phi^\dag \Phi \big]^3 \\
  &\  - \bigg[
     \frac{\bar c_{\sss u}}{v^2} y_u     \Phi^\dag \Phi\ \Phi^\dag\cdot{\bar Q}_L u_R
   + \frac{\bar c_{\sss d}}{v^2} y_d     \Phi^\dag \Phi\ \Phi {\bar Q}_L d_R
   + \frac{\bar c_{\sss l}}{v^2} y_\ell\ \Phi^\dag \Phi\ \Phi {\bar L}_L e_R
   + {\rm h.c.} \bigg] \\
  &\
  + \frac{i g\ \bar c_{\sss W}}{\mW^2} \big[ \Phi^\dag T_{2k} \overleftrightarrow{D}^\mu \Phi \big]  D^\nu  W_{\mu \nu}^k
  + \frac{i g'\ \bar c_{\sss B}}{2 \mW^2} \big[\Phi^\dag \overleftrightarrow{D}^\mu \Phi \big] \partial^\nu  B_{\mu \nu} \\
  &\
  + \frac{2 i g\ \bar c_{\sss HW}}{\mW^2} \big[D^\mu \Phi^\dag T_{2k} D^\nu \Phi\big] W_{\mu \nu}^k
  + \frac{i g'\ \bar c_{\sss HB}}{\mW^2}  \big[D^\mu \Phi^\dag D^\nu \Phi\big] B_{\mu \nu} \\
  &\
  +\frac{g'^2\ \bar c_{\sss \gamma}}{\mW^2} \Phi^\dag \Phi B_{\mu\nu} B^{\mu\nu}
   +\frac{g_s^2\ \bar  c_{\sss g}}{\mW^2} \Phi^\dag \Phi G_{\mu\nu}^a G_a^{\mu\nu}\ ,
\esp\ee
where the Wilson coefficients $\bar c$ are free parameters, $\lambda$ stands for the Higgs quartic coupling
and $y_u$, $y_d$ and $y_\ell$ are the $3\times 3$ Yukawa coupling matrices in flavor space
(all flavor indices are understood for clarity). In this expression, we also denote
the $U(1)_Y$, $SU(2)_L$ and $SU(3)_c$ coupling constants by $g'$, $g$ and $g_s$,
whereas the generators of $SU(2)$ in the fundamental representation are given by
$T_{2k} = \sigma_k/2$, $\sigma_k$ being the Pauli matrices.
Additionally, we have introduced
the Hermitian derivative operators ${\overleftrightarrow D}_\mu$ defined as
\be
  \Phi^\dag {\overleftrightarrow D}_\mu \Phi = 
    \Phi^\dag D^\mu \Phi - D_\mu\Phi^\dag \Phi \ ,
\ee
and the $SU(2)$ invariant products
\be
  Q_L\cdot\Phi = \epsilon_{ij}\ Q_L^i\ \Phi^j
  \quad\text{and}\quad
  \Phi^\dag\cdot \bar Q_L = \epsilon^{ij}\ \Phi^\dag_i\ \bar Q_{Lj} \ ,
\ee
the rank-two antisymmetric tensors being defined by $\epsilon_{12}=1$ and
$\epsilon^{12}=-1$. Finally,
our conventions for the gauge-covariant derivatives and the gauge field strength tensors are
\be\bsp
  B_{\mu\nu} =&\ \partial_\mu B_\nu - \partial_\nu B_\mu \ ,\\
  W^k_{\mu\nu} =&\ \partial_\mu W^k_\nu - \partial_\nu W^k_\mu + g \epsilon_{ij}{}^k \ W^i_\mu W^j_\nu\ ,\\
  G^a_{\mu\nu} =&\ \partial_\mu G^a_\nu - \partial_\nu G^a_\mu + g_s f_{bc}{}^a\ G^b_\mu G^c_\nu\ ,\\
  D_\rho W^k_{\mu\nu} = &\ \partial_\mu\partial_\rho W^k_\nu - \partial_\nu\partial_\rho W^k_\mu +
    g \epsilon_{ij}{}^k \partial_\rho\big[W_\mu^i W_\nu^j\big] +
    g \epsilon_{ij}{}^k W_\rho^i \big[\partial_\mu W_\nu^j - \partial_\nu W_\mu^j\big]\\ &\quad + 
    g^2  W_{\rho i}\big[W_\nu^i W_\mu^k - W_\mu^i W_\nu^k\big] \ , \\
  D_\mu\Phi =&\ \partial_\mu \Phi - \frac12 i g' B_\mu \Phi -  i g T_{2k} W_\mu^k \Phi \ , 
\esp\label{eq:covder}\ee
$\epsilon_{ij}{}^k$ and $f_{ab}{}^c$ being the structure constants of $SU(2)$ and
$SU(3)$.
The Lagrangian of Eq.~\eqref{eq:silh} can be supplemented by extra $CP$-violating operators,
\be\label{eq:silhCPodd}\bsp
  {\cal L}_{CP} = &\
    \frac{i g\ \tilde c_{\sss HW}}{\mW^2}  D^\mu \Phi^\dag T_{2k} D^\nu \Phi {\widetilde W}_{\mu \nu}^k
  + \frac{i g'\ \tilde c_{\sss HB}}{\mW^2} D^\mu \Phi^\dag D^\nu \Phi {\widetilde B}_{\mu \nu}
  + \frac{g'^2\  \tilde c_{\sss \gamma}}{\mW^2} \Phi^\dag \Phi B_{\mu\nu} {\widetilde B}^{\mu\nu}\\
 &\
  +\!  \frac{g_s^2\ \tilde c_{\sss g}}{\mW^2}      \Phi^\dag \Phi G_{\mu\nu}^a {\widetilde G}^{\mu\nu}_a
  \!+\!  \frac{g^3\ \tilde c_{\sss 3W}}{\mW^2} \epsilon_{ijk} W_{\mu\nu}^i W^\nu{}^j_\rho {\widetilde W}^{\rho\mu k}
  \!+\!  \frac{g_s^3\ \tilde c_{\sss 3G}}{\mW^2} f_{abc} G_{\mu\nu}^a G^\nu{}^b_\rho {\widetilde G}^{\rho\mu c} \ ,
\esp\ee
where the dual field strength tensors are defined by
\be
  \widetilde B_{\mu\nu} = \frac12 \epsilon_{\mu\nu\rho\sigma} B^{\rho\sigma} \ , \quad
  \widetilde W_{\mu\nu}^k = \frac12 \epsilon_{\mu\nu\rho\sigma} W^{\rho\sigma k} \ , \quad
  \widetilde G_{\mu\nu}^a = \frac12 \epsilon_{\mu\nu\rho\sigma} G^{\rho\sigma a} \ .
\ee

The third term in Eq.~\eqref{eq:effL} contains interactions between two Higgs fields and a
pair of quarks or leptons,
\be\label{eq:lf1}\bsp
  {\cal L}_{F_1} = &\ 
    \frac{i \bar c_{\sss HQ}}{v^2}  \big[\bar Q_L \gamma^\mu Q_L\big] \big[ \Phi^\dag{\overleftrightarrow D}_\mu \Phi\big]
  + \frac{4 i \bar c'_{\sss HQ}}{v^2} \big[\bar Q_L \gamma^\mu T_{2k} Q_L\big]  \big[\Phi^\dag T^k_2 {\overleftrightarrow D}_\mu \Phi\big] \\
  &\
  + \frac{i \bar c_{\sss Hu}}{v^2} \big[\bar u_R \gamma^\mu u_R\big]  \big[ \Phi^\dag{\overleftrightarrow D}_\mu \Phi\big]
  + \frac{i \bar c_{\sss Hd}}{v^2} \big[\bar d_R \gamma^\mu d_R\big]  \big[ \Phi^\dag{\overleftrightarrow D}_\mu \Phi\big]\\
  &\  -\bigg[
    \frac{i \bar c_{\sss Hud}}{v^2} \big[\bar u_R \gamma^\mu d_R\big]  \big[ \Phi \cdot {\overleftrightarrow D}_\mu \Phi\big]
   + {\rm h.c.} \bigg] \\
  &\ +
    \frac{i \bar c_{\sss HL}}{v^2}  \big[\bar L_L \gamma^\mu L_L\big] \big[ \Phi^\dag{\overleftrightarrow D}_\mu \Phi\big]
  + \frac{4 i \bar c'_{\sss HL}}{v^2} \big[\bar L_L \gamma^\mu T_{2k} L_L\big]  \big[\Phi^\dag T^k_2 {\overleftrightarrow D}_\mu \Phi\big] \\
  &\
  + \frac{i \bar c_{\sss He}}{v^2} \big[\bar e_R \gamma^\mu e_R\big]  \big[ \Phi^\dag{\overleftrightarrow D}_\mu \Phi\big] \ ,
\esp\ee
whilst the fourth term of this Lagrangian addresses the interactions of a quark or lepton pair and one single Higgs field
and a gauge boson,
\be\label{eq:lf2}\bsp
  {\cal L}_{F_2} = &\  \bigg[
  - \frac{2 g'\ \bar c_{\sss uB}}{\mW^2}  y_u\ \Phi^\dag \cdot {\bar Q}_L \gamma^{\mu\nu} u_R \  B_{\mu\nu}
  - \frac{4 g\ \bar c_{\sss uW}}{\mW^2}   y_u\ \Phi^\dag \cdot \big({\bar Q}_L T_{2k}\big) \gamma^{\mu\nu} u_R  \ W_{\mu\nu}^k\\
  &\quad
  - \frac{4 g_s\ \bar c_{\sss uG}}{\mW^2} y_u\ \Phi^\dag \cdot {\bar Q}_L \gamma^{\mu\nu} T_a u_R G_{\mu\nu}^a 
  + \frac{2 g'\ \bar c_{\sss dB}}{\mW^2}  y_d\ \Phi {\bar Q}_L \gamma^{\mu\nu} d_R \  B_{\mu\nu}\\
 &\quad
  + \frac{4 g\ \bar c_{\sss dW}}{\mW^2}   y_d\ \Phi \big({\bar Q}_L T_{2k}\big) \gamma^{\mu\nu} d_R  \ W_{\mu\nu}^k
  + \frac{4 g_s\ \bar c_{\sss dG}}{\mW^2} y_d\ \Phi {\bar Q}_L \gamma^{\mu\nu} T_a d_R G_{\mu\nu}^a \\
 &\quad
  + \frac{2 g'\ \bar c_{\sss eB}}{\mW^2}  y_\ell\ \Phi {\bar L}_L \gamma^{\mu\nu} e_R \  B_{\mu\nu}
  + \frac{4 g\ \bar c_{\sss eW}}{\mW^2}   y_\ell\ \Phi \big({\bar L}_L T_{2k}\big) \gamma^{\mu\nu} e_R  \ W_{\mu\nu}^k 
 +  {\rm h.c.} \bigg]\ .
\esp\ee
In this expression, the matrices $T_a$ are the generators of the $SU(3)$ group in the fundamental representation
and the $\gamma^{\mu\nu}$ quantities, defined by
\be
  \gamma^{\mu\nu} = \frac{i}{4} \Big[\gamma^\mu,\gamma^\nu\Big] \ ,
\ee
are the generators of the Lorentz algebra in the (four-component) spinorial representation.
In the most general case, the Wilson coefficients $\bar c_i$ related to the fermionic operators included in
the Lagrangians ${\cal L}_{F_i}$ are tensors in flavor space and complex quantities.

The last term of Eq.~\eqref{eq:effL} refers to operators not directly connected to Higgs physics, but that may
be important as affecting the gauge sector and possibly modifying the
gauge boson self-energies and self-interactions,
\be\bsp
  {\cal L}_{G} = &\
    \frac{g^3\ \bar c_{\sss 3W}}{\mW^2} \epsilon_{ijk} W_{\mu\nu}^i W^\nu{}^j_\rho W^{\rho\mu k}
  + \frac{g_s^3\ \bar c_{\sss 3G}}{\mW^2} f_{abc} G_{\mu\nu}^a G^\nu{}^b_\rho G^{\rho\mu c}
  + \frac{\bar c_{\sss 2W}}{\mW^2} D^\mu W_{\mu\nu}^k D_\rho W^{\rho\nu}_k \\
 &\
  + \frac{\bar c_{\sss 2B}}{\mW^2} \partial^\mu B_{\mu\nu} \partial_\rho B^{\rho\nu}
  + \frac{\bar c_{\sss 2G}}{\mW^2} D^\mu G_{\mu\nu}^a D_\rho G^{\rho\nu}_a\ ,
\esp\label{eq:lagG}\ee
with
\be\bsp
  D_\rho G^a_{\mu\nu} = &\ \partial_\mu\partial_\rho G^a_\nu - \partial_\nu\partial_\rho G^a_\mu +
    g_s f_{bc}{}^a \partial_\rho\big[G_\mu^b G_\nu^c\big] +
    g_s f_{bc}{}^a G_\rho^b \big[\partial_\mu G_\nu^c - \partial_\nu G_\mu^c\big]\\ &\quad + 
    g_s^2  G_{\rho b}\big[G_\nu^b G_\mu^a - G_\mu^b G_\nu^a\big] \ ,
\esp\ee
recalling that $D_\mu W^{\nu\rho}_k$ has been defined in Eq.~\eqref{eq:covder}.

Finally, 22 independent
baryon and lepton number conserving four-fermion operators are also allowed by gauge invariance. Since they
have no effects on Higgs physics, at least at the leading order and in the context
of the LHC phenomenology, we omit them from the present manuscript, as already
above-mentioned. Moreover, as indicated in
Ref.~\cite{Contino:2013kra}, two of the 39 operators that have been introduced are redundant and can be removed through
\be\bsp
  {\cal O}_{\sss W} =&\ -2 {\cal O}_{\sss H} + \frac{4}{v^2} \Phi^\dag \Phi D^\mu\Phi^\dag D_\mu\Phi + {\cal O}'_{\sss HQ} + 
    {\cal O}'_{\sss HL} \ , \\
  {\cal O}_{\sss B} =&\ 2 \tan^2\theta_{\sss W} \Big[ \sum_\psi Y_\psi {\cal O}_{\sss H\psi} - {\cal O}_{\sss T} \Big] \ ,
\esp\ee
where we sum over the whole chiral content of the theory and $\theta_{\sss W}$ stands for the weak mixing angle
(see Eq.~\eqref{eq:weakmix1} and Eq.~\eqref{eq:weakmix2} in Section~\ref{sec:massbasis}).
We however include them in our effective field
theory description as according to the specific effect of interest, one choice for an operator basis
may be more suitable than another.

The full set of interactions generated by the 39 operators
included in the Lagrangian ${\cal L}$ of
Eq.~\eqref{eq:effL} have been implemented in a {\sc FeynRules}~\cite{Christensen:2008py,
Alloul:2013bka} model file
that can be either downloaded from

\verb+http://feynrules.irmp.ucl.ac.be/wiki/HEL+

\noindent or obtained from the authors upon request. Employing the various
interfaces linking {\sc FeynRules} to Monte Carlo event generators,
the effects of those operators can be further studied at colliders. Care must
however be
taken with the choice of the Monte Carlo generator to employ for
this purpose, as most of them
have strong requirements on the allowed Lorentz structures for the vertices\footnote{The
interfaces take care of discarding each vertex that is not compliant
with the relevant program.}. To avoid such limitations and be capable of
probing any of the 39 considered operators, we make use, for the few examples of
Section~\ref{sec:pheno}, of the {\sc MadGraph}~5
program~\cite{Alwall:2011uj} linked to the UFO~\cite{Degrande:2011ua}
version of the model presented above.

\subsection{The most general effective Higgs Lagrangian in the mass eigenbasis}
\label{sec:massbasis}
After electroweak symmetry breaking to electromagnetism, \ie, 
when the Higgs field acquires its vacuum expectation value,
the gauge eigenstates introduced in Eq.~\eqref{eq:gaugecontent} mix to the physical massive
$W$-boson and $Z$-boson as well as to the photon $A$ which remains massless,
\be\label{eq:weakmix1}\bsp
  W_\mu^\pm =&\ {1\over \sqrt{2}}  (W_\mu^1 \mp i W^2_\mu)\ ,\\
  \bpm Z_\mu\\ A_\mu \epm =&\
    \bpm \cos\theta_{\sss W} & -\sin\theta_{\sss W} \\ \sin\theta_{\sss W} & \phantom{-}\cos\theta_{\sss W}\epm 
    \bpm W_\mu^3 \\ B_\mu\epm \equiv \bpm \cW & -\sW \\ \sW & \phantom{-}\cW\epm \bpm W_\mu^3 \\ B_\mu\epm \ .
\esp\ee
In this equation, we have introduced
the weak mixing angle at tree-level whose sine and cosine, noted as
$\sW$ and $\cW$, are defined with respect to the
hypercharge ($g^\prime$), weak ($g$) and electromagnetic ($e$)
coupling constants,
\be\label{eq:weakmix2}
  g' \cW = g \sW = e \ .
\ee

First, it can be observed  that
some of the new physics operators induce a modification of
the kinetic terms of the gauge and Higgs bosons, once the neutral
component of the Higgs doublet gets its vacuum expectation value $v$.
For instance,
the ${\cal O}_G$ operator of Eq.~\eqref{eq:silh}
leads to a variation of the gluon field strength tensor squared term
given by
\be
  {\cal O}_G = \frac{g_s^2\ \bar c_{\sss g}}{\mW^2} \Phi^\dag \Phi G_{\mu\nu}^a G_a^{\mu\nu} \qquad \Rightarrow\qquad
    -\frac{g_s^2\ \bar c_{\sss g} v^2}{2 \mW^2} G_{\mu\nu}^a G_a^{\mu\nu} \ ,
\ee
and the ${\cal O}_H$ operator implies an additional contribution
to the Higgs boson $h$ kinetic term,
\be
  {\cal O}_H =  \frac{\bar c_{\sss H}}{2 v^2} \partial^\mu\big[\Phi^\dag \Phi\big] \partial_\mu \big[ \Phi^\dagger \Phi \big]
    \qquad\Rightarrow\qquad \frac{\bar c_{\sss H}}{2} \partial^\mu h \partial_\mu h \ .
\ee
Appropriate field redefinitions (at the first order
in the effective couplings) are then required
in order to bring back the various kinetic
terms to their canonical forms, and read
\be\bsp
 h \to &\ h \bigg[ 1-\frac12 c_{\sss H}\bigg] \ ,\\
 g_\mu^a \to &\ g_\mu^a \bigg[1 + \frac{\bar c_{\sss g} g_s^2 v^2}{\mW^2}\bigg] \ , \\
 Z_\mu \to &\ Z_\mu \bigg[1 + \frac{\bar c_{\sss \gamma} g^2 \sW^4 v^2}{\cW^2 \mW^2}\bigg] \ ,\\
 A_\mu \to &\ A_\mu \bigg[1 + \frac{\bar c_{\sss \gamma} e^2 v^2}{\mW^2}\bigg] - Z_\mu \bigg[\frac{2 \bar c_{\sss \gamma} \sW e^2 v^2}{\cW \mW^2}\bigg] \ .
\esp\label{eq:fieldred}\ee
While in our operator basis convention, the mass of the
$W$-boson is agnostic
of any new physics effect, there is an impact on both the $Z$-boson and
Higgs-boson masses,
\be\bsp
  m_{\sss Z}^2 =&\ \frac{g^2 v^2}{4 c_w^2}\Big[ 1- \bar c_{\sss T} +
    \frac{8 \bar c_{\sss\gamma} \sW^4}{\cW^2}\Big] \ , \\
  m_{\sss H}^2 = &\ 2 v^2 \lambda \Big[1+\frac{13}{8} \bar c_{\sss 6} -
     \bar c_{\sss H}\Big] \ .
\esp\ee
In this last relation, we have made use of the minimization condition
of the Higgs potential that reads, again in the first order
in the effective couplings,
\be
  v^2 = \frac{\mu^2}{\lambda} \Big[ 1-\frac12 \bar c_{\sss 6}\Big] \ .
\ee
These considerations and simplicity have motivated us
to choose the electroweak input parameters that are used in the
{\sc FeynRules} model file. They consist in the mass of the $W$-boson
$\mW$, the Fermi constant $G_{\sss F}$, the
electromagnetic coupling constant $\alpha$, as well as
the Higgs-boson mass
$m_{\sss H}$. The field redefinitions of Eq.~\eqref{eq:fieldred},
the replacement of the internal parameters in terms of the
external ones and the removal of any term of higher order
that is possibly arising from the procedure sketched above
have been fully automated.

One could have implemented the effective Lagrangian of Section~\ref{sec:HELgauge} directly
in the mass basis. This choice, adopted for instance in Ref.~\cite{Artoisenet:2013puc},
is in general clearer when one aims to properly
estimate new physics effects that can be hinted for by a given
physical observable. Several of the effective operators introduced
in the previous section can indeed give rise to the same interaction
in the mass basis, with the same Lorentz
structure, so that it is not straightforward, if not impossible,
to map a single measurement to a single
operator.
Moreover, a Lagrangian expressed in
the mass basis also offers an easy way to be generalized by including
appropriate form factors to model new physics effects. We consequently
dedicate this section to a proper comparison of the two approaches,
linking the effective Lagrangian of Section~\ref{sec:HELgauge} that is expressed entirely
in the gauge basis to its counterpart as read off in terms of
mass-eigenstates after electroweak symmetry
breaking.

We now turn to the investigation of the interaction terms
of the Lagrangian.
Focusing on the terms related to the Higgs sector one writes
\be
  {\cal L}_{\rm Higgs} = {\cal L}^{(3)} + {\cal L}^{(4)} + {\cal L}^{(5)} + {\cal L}^{(6)} \ ,
\ee
where ${\cal L}_i$ denotes the set of $i$-point interactions involving at least one Higgs boson.
For the sake of the example, we only work out explicitly the structure of the interactions
which are the more relevant
for the phenomenology of the Higgs sector at the LHC, namely the three-point and four-point
interactions involving at least one Higgs field.
Vertices involving a higher number of external legs are in general related to processes
associated with smaller cross sections, making them difficult to probe.
We therefore refer to the {\sc FeynRules} implementation for their explicit form in the mass basis.

\renewcommand{\arraystretch}{1.5}
\begin{table}
  \center
  \begin{tabular}{l| l| l}
  \hline\hline
    Eq.~\eqref{eq:massbasis} & 
    Ref.~\cite{Artoisenet:2013puc}&
    Section~\ref{sec:HELgauge}\\
  \hline\hline
    $g_{\sss hgg}$        & $c_\alpha \kappa_{\sss Hgg} g_{\sss Hgg}$ & $g_{\sss H}
    - \frac{4 \bar c_{\sss g} g_s^2 v}{\mW^2}$  \\
    $\tilde g_{\sss hgg}$ & $s_\alpha \kappa_{\sss Agg} g_{\sss Agg}$ & $ -\frac{4 \tilde c_{\sss g} g_s^2 v}{\mW^2}$ \\
    $g_{\sss h\gamma\gamma}$ & $c_\alpha \kappa_{\sss H\gamma\gamma} g_{\sss H\gamma\gamma}$ & $a_{\sss H}
       - \frac{8 g \bar c_{\sss \gamma} \sW^2}{\mW}$ \\
    $\tilde g_{\sss h\gamma\gamma}$ & $s_\alpha \kappa_{\sss A\gamma\gamma} g_{\sss A\gamma\gamma}$ &
     $ -\frac{8 g \tilde c_{\sss \gamma} \sW^2}{\mW}$ \\
    $g^{(1)}_{\sss hzz}$ & $\frac{1}{\Lambda} c_\alpha \kappa_{\sss HZZ}$ & 
       $\frac{2 g}{\cW^2 \mW} \Big[ \bar c_{\sss HB} \sW^2 - 4 \bar c_{\sss \gamma} \sW^4 + \cW^2 \bar c_{\sss HW}\Big]$ \\
    $\tilde g_{\sss hzz}$ & $\frac{1}{\Lambda} s_\alpha \kappa_{\sss AZZ}$ & 
       $\frac{2 g}{\cW^2 \mW} \Big[ \tilde c_{\sss HB} \sW^2 - 4 \tilde c_{\sss \gamma} \sW^4 + \cW^2 \tilde c_{\sss HW}\Big]$ \\
    $g^{(2)}_{\sss hzz}$ &$\frac{1}{\Lambda} c_\alpha \kappa_{\sss H\partial Z}$ &
        $\frac{g}{\cW^2 \mW} \Big[(\bar c_{\sss HW} +\bar c_{\sss W}) \cW^2  + (\bar c_{\sss B} + \bar c_{\sss HB}) \sW^2 \Big]$\\
    $g^{(3)}_{\sss hzz}$  & $c_\alpha \kappa_{\rm SM} g_{\sss HZZ}$& $\frac{g \mW}{\cW^2} \Big[ 1 -\frac12 \bar c_{\sss H} - 2 \bar c_{\sss T}
          +8 \bar c_{\sss\gamma} \frac{\sW^4}{\cW^2}  \Big]$ \\
    $g^{(1)}_{\sss haz}$ & $c_\alpha \kappa_{\sss HZ\gamma} g_{\sss HZ\gamma}$ &
          $\frac{g \sW}{\cW \mW} \Big[  \bar c_{\sss HW} - \bar c_{\sss HB} + 8 \bar c_{\sss \gamma} \sW^2\Big]$ \\
    $\tilde g_{\sss haz}$ & $s_\alpha \kappa_{\sss AZ\gamma} g_{\sss AZ\gamma}$ &
         $\frac{g \sW}{\cW \mW} \Big[  \tilde c_{\sss HW} - \tilde c_{\sss HB} + 8 \tilde c_{\sss \gamma} \sW^2\Big]$ \\
    $g^{(2)}_{\sss haz}$ & $\frac{1}{\Lambda} c_\alpha \kappa_{\sss H\partial \gamma}$ &
      $\frac{g \sW}{\cW \mW} \Big[  \bar c_{\sss HW} - \bar c_{\sss HB} - \bar c_{\sss B} + \bar c_{\sss W}\Big]$ \\
    $g^{(1)}_{\sss hww}$ &$\frac{1}{\Lambda} c_\alpha \kappa_{\sss HWW}$ &  $\frac{2 g}{\mW} \bar c_{\sss HW}$ \\
    $\tilde g_{\sss hww}$ &$\frac{1}{\Lambda} s_\alpha \kappa_{\sss AWW}$ & $\frac{2 g}{\mW} \tilde c_{\sss HW}$ \\
    $g^{(2)}_{\sss hww}$ & $\frac{1}{\Lambda} c_\alpha \kappa_{\sss H\partial W}$ &
      $\frac{g}{\mW} \Big[ \bar c_{\sss W} + \bar c_{\sss HW} \Big]$ \\
  \end{tabular}
  \caption{Coupling strengths of the interactions of a Higgs boson with a vector boson pair.
    We present the relations between the Lagrangian parameters introduced in Eq.~\eqref{eq:massbasis} (first column), where
    the Lagrangian is expressed in the mass basis, and those associated with the operators of Section~\ref{sec:HELgauge}
    expressed in the gauge basis (third column).  The Standard Model contributions to the Higgs boson
    to two photons (gluons) vertex
    $a_{\sss H}$ ($g_{\sss H}$) have been explicitly indicated. We relate these parameters to those employed in
    the Lagrangian description of Ref.~\cite{Artoisenet:2013puc} in the second column of the table.}
  \label{tab:paramshvv}
\end{table}
\renewcommand{\arraystretch}{1}

\renewcommand{\arraystretch}{1.5}
\begin{table}
  \center
  \begin{tabular}{l| l }
  \hline\hline
    Eq.~\eqref{eq:massbasis} -
    Eq.~\eqref{eq:massbasis4} & 
    Section~\ref{sec:HELgauge}\\
  \hline \hline
  $g_{\sss hhh}^{(1)}$   & $1 + \frac78 \bar c_{\sss 6} - \frac12 \bar c_{\sss H}$ \\
  $g_{\sss hhh}^{(2)}$   & $\frac{g}{\mW} \bar c_{\sss H}$ \\
  $g_{\sss hhhh}^{(1)}$ & $1 + \frac{47}{8} \bar c_{\sss 6} - \bar c_{\sss H}$ \\
  $g_{\sss hhhh}^{(2)}$ & $\frac{g^2}{4 \mW^2} \bar c_{\sss H}$ \\
  \end{tabular}
  \caption{Multiple Higgs interactions.
    We present the relations between the Lagrangian parameters introduced in Eq.~\eqref{eq:massbasis}
    and Eq.~\eqref{eq:massbasis4}, where
    the Lagrangian is expressed in the mass basis, and those associated with the operators
    of Section~\ref{sec:HELgauge} expressed in the gauge basis.}
  \label{tab:paramshhh}
\end{table}
\renewcommand{\arraystretch}{1}

\renewcommand{\arraystretch}{1.45}
\begin{table}
  \center
  \begin{tabular}{l| c| l}
  \hline\hline
    Eq.~\eqref{eq:massbasis}  - Eq.~\eqref{eq:massbasis4}& 
    Ref.~\cite{Artoisenet:2013puc}&
    Section~\ref{sec:HELgauge}\\
  \hline\hline
    $\tilde y_u$    & $c_\alpha \kappa_{\sss Huu} g_{\sss Huu}$ & $y_u \Big[1 -\frac12 \bar c_{\sss H} + \frac32 \bar c_{\sss u}\Big]$ \\
    $\tilde y_d$    & $c_\alpha \kappa_{\sss Hdd} g_{\sss Hdd}$ & $y_d \Big[1  -\frac12 \bar c_{\sss H} + \frac32 \bar c_{\sss d}\Big]$ \\
    $\tilde y_\ell$ & $c_\alpha \kappa_{\sss H\ell\ell} g_{\sss H\ell\ell}$ & $y_\ell \Big[1  -\frac12 \bar c_{\sss H} + \frac32 \bar c_{\sss \ell}\Big]$\\
    $\bar y_u$    & - & $\frac{y_u}{v} \frac32 \bar c_{\sss u}$ \\
    $\bar y_d$    & - & $\frac{y_d}{v} \frac32 \bar c_{\sss d}$ \\
    $\bar y_\ell$ & - & $\frac{y_\ell}{v} \frac32 \bar c_{\sss \ell}$\\
    $\Big\{g_{\sss hzuu}^{(L)}, g_{\sss hzuu}^{(R)}\Big\}$ & - & $\frac{g}{\cW v} \Big\{ \bar c_{\sss HQ} - \bar c'_{\sss HQ}, 
        \bar c_{\sss Hu} \Big\}$\\
    $\Big\{g_{\sss hzdd}^{(L)}, g_{\sss hzdd}^{(R)}\Big\}$ & - & $\frac{g}{\cW v} \Big\{ \bar c_{\sss HQ} + \bar c'_{\sss HQ}, 
        \bar c_{\sss Hd} \Big\}$\\
    $\Big\{g_{\sss hz\ell\ell}^{(L)}, g_{\sss hz\ell\ell}^{(R)}\Big\}$ & - &
      $\frac{g}{\cW v} \Big\{ \bar c_{\sss HL} + \bar c'_{\sss HL}, \frac12 \bar c_{\sss e} \Big\}$\\
    $g_{\sss hz\nu\nu}$ & - & $\frac{g}{\cW v} \Big[\bar c_{\sss HL} - \bar c'_{\sss HL} \Big]$\\
    $\Big\{g_{\sss hwud}^{(L)}, g_{\sss hwud}^{(R)}\Big\}$ & - & $\frac{\sqrt{2} g}{v} \Big\{ \bar c'_{\sss HQ} V^{\sss \rm CKM},
        \bar c_{\sss Hud} \Big\}$\\
    $g_{\sss hw\nu\ell}$ & - & $\frac{\sqrt{2} g}{v} \bar c'_{\sss HL}$\\
    $g_{\sss h\gamma uu}^{(\partial)}$ & - & $\frac{\sqrt{2} g \sW}{\mW^2} y_u \Big[ \bar c_{\sss uB} + \bar c_{\sss uW}\Big]$\\
    $g_{\sss h\gamma dd}^{(\partial)}$ & - & $\frac{\sqrt{2} g \sW}{\mW^2} y_d \Big[ \bar c_{\sss dB} - \bar c_{\sss dW}\Big]$\\
    $g_{\sss h\gamma \ell\ell}^{(\partial)}$ & - & $\frac{\sqrt{2} g \sW}{\mW^2} y_\ell \Big[ \bar c_{\sss eB} - \bar c_{\sss eW}\Big]$\\
    $g_{\sss hzuu}^{(\partial)}$ & - & $\frac{\sqrt{2} g}{\cW \mW^2} y_u \Big[ \bar c_{\sss uW} \cW^2 - \bar c_{\sss uB} \sW^2 \Big]$\\
    $g_{\sss hzdd}^{(\partial)}$ & - & $\frac{\sqrt{2} g}{\cW \mW^2} y_d \Big[ -\bar c_{\sss dW} \cW^2 - \bar c_{\sss dB} \sW^2 \Big]$\\
    $g_{\sss hz\ell\ell}^{(\partial)}$ & - & $\frac{\sqrt{2} g}{\cW \mW^2} y_\ell \Big[ -\bar c_{\sss eW} \cW^2 - \bar c_{\sss eB} \sW^2 \Big]$\\
    $\Big\{g_{\sss hwud}^{(\partial L)}, g_{\sss hwud}^{(\partial R)}\Big\}$ & - &
       $\frac{2 g}{\mW^2} \Big\{y^\dag_u V^{\sss\rm CKM} \bar c_{\sss uW},  V^{\sss \rm CKM} y_d \bar c_{\sss dW} \Big\}$\\
    $g_{\sss hw\nu\ell}^{(\partial)}$ & - &
       $\frac{2 g}{\mW^2} y_\ell \bar c_{\sss eW}$\\
    $g_{\sss hguu}^{(\partial)}$ & - & $\frac{2 \sqrt{2} g_s}{\mW^2} y_u \bar c_{\sss uG}$\\
    $g_{\sss hgdd}^{(\partial)}$ & - & $\frac{2 \sqrt{2} g_s}{\mW^2} y_d \bar c_{\sss dG}$\\
  \end{tabular}
  \caption{Coupling strengths of the interactions of one or several Higgs boson(s)
    with a fermion pair and possibly an additional gauge boson.
    We present the relations between the Lagrangian parameters introduced in Eq.~\eqref{eq:massbasis} (first column), where
    the Lagrangian is expressed in the mass basis, and those associated with the operators of Section~\ref{sec:HELgauge}
    expressed in the gauge basis (third column). In our notations, $V^{\sss \rm CKM}$ denotes
    the CKM mixing matrix. Concerning the three-point interactions,
    we relate our parameters to those of
    the Lagrangian description of Ref.~\cite{Artoisenet:2013puc} in the second column of the table.}
  \label{tab:paramshff}
\end{table}
\renewcommand{\arraystretch}{1}

In the unitarity gauge and in the mass basis, the ${\cal L}_3$ Lagrangian reads
\be\bsp
  {\cal L}_3 = &\ 
    - \frac{m_{\sss H}^2}{2 v} g^{(1)}_{\sss hhh}h^3 + \frac{1}{2} g^{(2)}_{\sss hhh} h\partial_\mu h \partial^\mu h
\\ &\ 
    - \frac{1}{4} g_{\sss hgg} G^a_{\mu\nu} G_a^{\mu\nu} h
    - \frac{1}{4} \tilde g_{\sss hgg} G^a_{\mu\nu} \tilde G^{\mu\nu} h
    - \frac{1}{4} g_{\sss h\gamma\gamma} F_{\mu\nu} F^{\mu\nu} h
    - \frac{1}{4} \tilde g_{\sss h\gamma\gamma} F_{\mu\nu} \tilde F^{\mu\nu} h
\\ &\ 
    - \frac{1}{4} g_{\sss hzz}^{(1)} Z_{\mu\nu} Z^{\mu\nu} h
    - g_{\sss hzz}^{(2)} Z_\nu \partial_\mu Z^{\mu\nu} h
    + \frac{1}{2} g_{\sss hzz}^{(3)} Z_\mu Z^\mu h
    - \frac{1}{4} \tilde g_{\sss hzz} Z_{\mu\nu} \tilde Z^{\mu\nu} h
\\ &\
    - \frac{1}{2} g_{\sss haz}^{(1)} Z_{\mu\nu} F^{\mu\nu} h
    - \frac{1}{2} \tilde g_{\sss haz} Z_{\mu\nu} \tilde F^{\mu\nu} h
    - g_{\sss haz}^{(2)} Z_\nu \partial_\mu F^{\mu\nu} h
    - \frac{1}{2} g_{\sss hww}^{(1)} W^{\mu\nu} W^\dag_{\mu\nu} h
\\ &\
    - \Big[g_{\sss hww}^{(2)} W^\nu \partial^\mu W^\dag_{\mu\nu} h + {\rm h.c.} \Big]
    +  g (1-\frac12 \bar c_{\sss H}) \mW W_\mu^\dag  W^\mu h
    - \frac{1}{2} \tilde g_{\sss hww} W^{\mu\nu} \tilde W^\dag_{\mu\nu} h
\\ &\
    - \bigg[ 
      \tilde y_u \frac{1}{\sqrt{2}} \big[{\bar u} P_R u\big] h +
      \tilde y_d \frac{1}{\sqrt{2}} \big[{\bar d} P_R d\big] h +
      \tilde y_\ell \frac{1}{\sqrt{2}} \big[{\bar \ell} P_R \ell\big] h
     + {\rm h.c.} \bigg] \ ,
\esp\label{eq:massbasis}\ee
where the flavor indices of the fermions are understood
and where we have introduced the $W$-boson, $Z$-boson and photon field strength tensors,
$W_{\mu\nu}$, $Z_{\mu\nu}$ and
$F_{\mu\nu}$.
As already mentioned above and in Ref.~\cite{Artoisenet:2013puc}, this form of Lagrangian
is sufficient to characterize all Higgs properties in a non-ambiguous
way. This contrasts with the initial set of operators in the gauge basis which leads
to additional structures that can be removed after several integrations by parts.
In addition to all the operators already included in the Lagrangian of
Ref.~\cite{Artoisenet:2013puc}, \ie, the
Higgs to diboson couplings defined in Table~\ref{tab:paramshvv}
and the Higgs to fermions interactions of Table~\ref{tab:paramshff}, we
also include the triple Higgs interactions
presented in Table~\ref{tab:paramshhh}. Furthermore, the tables also contain a translation
dictionary linking our free parameters to the general Lagrangian employed
in Ref.~\cite{Artoisenet:2013puc}.

The set of four-point interactions involving one or several Higgs fields is deduced from
\be\bsp
  {\cal L}_4 = &\
    - \frac{m_{\sss H}^2}{8 v^2} g_{\sss hhhh}^{(1)} h^4
    + \frac12 g_{\sss hhhh}^{(2)} h^2 \partial_\mu h\partial^\mu h
    - \frac{1}{8} g_{\sss hhgg} G^a_{\mu\nu} G_a^{\mu\nu} h^2
    - \frac{1}{8} \tilde g_{\sss hhgg} G^a_{\mu\nu} \tilde G_a^{\mu\nu} h^2
\\ &\ 
    -\! \frac18 g_{\sss hh\gamma\gamma} F_{\mu\nu} F^{\mu\nu} h^2
    \!-\! \frac18 \tilde g_{\sss hh\gamma\gamma} F_{\mu\nu} \tilde F^{\mu\nu} h^2
    \!-\! \frac18 g_{\sss hhzz}^{(1)} Z_{\mu\nu} Z^{\mu\nu} h^2
    \!-\! \frac18 \tilde g_{\sss hhzz} Z_{\mu\nu} \tilde Z^{\mu\nu} h^2
\\ &\
    - \frac12 g_{\sss hhzz}^{(2)} Z_\nu \partial_\mu Z^{\mu\nu} h^2
    + \frac14 g_{\sss hhzz}^{(3)} Z_\mu Z^\mu h^2
    - \frac{1}{4} g_{\sss hhaz}^{(1)} Z_{\mu\nu} F^{\mu\nu} h^2
    - \frac{1}{4} \tilde g_{\sss hhaz} Z_{\mu\nu} \tilde F^{\mu\nu} h^2
\\ &\
    - \frac12 g_{\sss hhaz}^{(2)} Z_\nu \partial_\mu F^{\mu\nu} h^2
    - \frac14 g_{\sss hhww}^{(1)} W^{\mu\nu} W^\dag_{\mu\nu} h^2
    - \frac14 \tilde g_{\sss hhww} W^{\mu\nu} \tilde W^\dag_{\mu\nu} h^2
\\ &\
    - \frac12 \Big[g_{\sss hhww}^{(2)} W^\nu \partial^\mu W^\dag_{\mu\nu} h^2 + {\rm h.c.} \Big]
    + \frac14 g^2 (1-\bar c_{\sss H})W_\mu^\dag  W^\mu h^2
    - i g_{\sss haww}^{(1)} F^{\mu\nu} W_\mu W^\dag_\nu h
\\ &\
    + \Big[i g_{\sss haww}^{(2)} W^{\mu\nu} A_\mu W^\dag_\nu h + {\rm h.c.} \Big]
    + i g_{\sss haww}^{(3)} A_\mu W_\nu W^\dag_\rho \big[ \eta^{\mu\rho}\partial^\nu h - \eta^{\mu\nu}\partial^\rho h\big] 
\\ &\
    + i \tilde g_{\sss haww}^{(1)} \tilde F^{\mu\nu} W_\mu W^\dag_\nu h
    + \Big[i \tilde g_{\sss haww}^{(2)} \tilde W^{\mu\nu} A_\mu W^\dag_\nu h + {\rm h.c.} \Big]
\\ &\
    - i g_{\sss hzww}^{(1)} Z^{\mu\nu} W_\mu W^\dag_\nu  h
    + \Big[i g_{\sss hzww}^{(2)} W^{\mu\nu} Z_\mu W^\dag_\nu h + {\rm h.c.} \Big]
\\ &\
    + i \tilde g_{\sss hzww}^{(1)} \tilde Z^{\mu\nu} W_\mu W^\dag_\nu  h
    - \Big[i \tilde g_{\sss hzww}^{(2)} \tilde W^{\mu\nu} Z_\mu W^\dag_\nu h + {\rm h.c.} \Big]
\\ &\
    - i g_{\sss hzww}^{(3)} Z_\mu W_\nu W^\dag_\rho \big[ \eta^{\mu\rho}\partial^\nu h - \eta^{\mu\nu}\partial^\rho h\big] 
\\ &\
    - \bigg[ 
      \bar y_u \frac{1}{\sqrt{2}} \big[{\bar u} P_R u\big] h^2 +
      \bar y_d \frac{1}{\sqrt{2}} \big[{\bar d} P_R d\big] h^2 +
      \bar y_\ell \frac{1}{\sqrt{2}} \big[{\bar \ell} P_R \ell\big] h^2
     + {\rm h.c.} \bigg] 
\\ &\
    - {\bar u} \gamma^\mu \Big[ g_{\sss hzuu}^{(L)} P_L +  g_{\sss hzuu}^{(R)} P_R \Big] u Z_\mu h
    - {\bar d} \gamma^\mu \Big[ g_{\sss hzdd}^{(L)} P_L +  g_{\sss hzdd}^{(R)} P_R \Big] d Z_\mu h
\\ &\
    - {\bar \ell} \gamma^\mu \Big[ g_{\sss hz\ell\ell}^{(L)} P_L +  g_{\sss hz\ell\ell}^{(R)} P_R \Big] \ell Z_\mu h
    - {\bar \nu} \gamma^\mu \Big[ g_{\sss hz\nu\nu} P_L \Big] \nu Z_\mu h
\\ &\
    -\bigg[
        {\bar u} \gamma^\mu \Big[ g_{\sss hwud}^{(L)} P_L +  g_{\sss hwud}^{(R)} P_R \Big] d W_\mu h
      + {\bar \nu} \gamma^\mu \Big[ g_{\sss hw\nu\ell} P_L \Big] \ell W_\mu h
     + {\rm h.c.} \bigg]
\\ &\
    -\bigg[
       g_{\sss h\gamma uu}^{(\partial)} \Big[ {\bar u} \gamma^{\mu\nu} P_R u \Big]
     + g_{\sss h\gamma dd}^{(\partial)} \Big[ {\bar d} \gamma^{\mu\nu} P_R d \Big]
     + g_{\sss h\gamma \ell\ell}^{(\partial)} \Big[ {\bar \ell} \gamma^{\mu\nu} P_R \ell \Big]
     + {\rm h.c.} \bigg]  F_{\mu\nu} h
\\ &\
    -\bigg[
       g_{\sss hzuu}^{(\partial)} \Big[ {\bar u} \gamma^{\mu\nu} P_R u \Big]
     + g_{\sss hzdd}^{(\partial)} \Big[ {\bar d} \gamma^{\mu\nu} P_R d \Big]
     + g_{\sss hz\ell\ell}^{(\partial)} \Big[ {\bar \ell} \gamma^{\mu\nu} P_R \ell \Big]
     + {\rm h.c.} \bigg]  Z_{\mu\nu} h
\\ &\
    -\bigg[
       {\bar u} \gamma^{\mu\nu} \Big[ g_{\sss hwud}^{(\partial L)} P_L + g_{\sss hwud}^{(\partial R)} P_R\Big] d W_{\mu\nu}
     + g_{\sss hw\nu\ell}^{(\partial)} {\bar \nu} \gamma^{\mu\nu} P_R \ell W_{\mu\nu}
     + {\rm h.c.} \bigg]  h
\\ &\
    -\bigg[
       g_{\sss hguu}^{(\partial)} \Big[ {\bar u} T_a \gamma^{\mu\nu} P_R u \Big]
     + g_{\sss hgdd}^{(\partial)} \Big[ {\bar d} T_a \gamma^{\mu\nu} P_R d \Big]
     + {\rm h.c.} \bigg]  G_{\mu\nu}^a h \ ,
\esp\label{eq:massbasis4}\ee
where the free parameters are given in the various tables of this section. Hence,
Table~\ref{tab:paramshff} contains the parameters related to the interactions among one
Higgs boson $h$, a fermion-antifermion pair
and possibly an additional gauge boson, while Table~\ref{tab:paramshhh} and Table~\ref{tab:params4hhvv}
are respectively dedicated to the Higgs boson self-interactions and to its interactions with vector bosons.
We recall that some of the interaction vertices among a single Higgs field and three gluons were already
included in the Lagrangian of Eq.~\eqref{eq:massbasis} through the gluon field strength tensor and its dual. Consequently,
these are omitted from Eq.~\eqref{eq:massbasis4}. Along the same line, we include in a gauge-invariant way higher-dimensional
Higgs boson and gluon vertices.

\renewcommand{\arraystretch}{1.5}
\begin{table}
  \center
  \begin{tabular}{l| l }
  \hline\hline
    Eq.~\eqref{eq:massbasis4} & 
    Section~\ref{sec:HELgauge}\\
  \hline \hline
    $g_{\sss hhgg}$ & $-\frac{4 \bar c_{\sss g} g_s^2}{\mW^2}$ \\
    $\tilde g_{\sss hhgg}$ & $-\frac{4 \tilde c_{\sss g} g_s^2}{\mW^2}$ \\
    $g_{\sss hh\gamma\gamma}$ & $-\frac{4 \bar c_{\sss \gamma} g^2 \sW^2}{\mW^2}$ \\
    $\Big\{\tilde g_{\sss hh\gamma\gamma}, \tilde g_{\sss hhzz}, \tilde g_{\sss hhaz}, \tilde g_{\sss hhww}\Big\}$ & 
      $\frac{g}{2 \mW} \Big\{ \tilde g_{\sss h\gamma\gamma}, \tilde g_{\sss hzz}, \tilde g_{\sss haz}, 
      \tilde g_{\sss hww} \Big\}$ \\
    $\Big\{g^{(1)}_{\sss hhzz}, g^{(2)}_{\sss hhzz}, g^{(1)}_{\sss hhaz}, g^{(2)}_{\sss hhaz}, 
       g^{(1)}_{\sss hhww}, g^{(2)}_{\sss hhww} \Big\}$ & 
       $\frac{g}{2 \mW} \Big\{g_{\sss hzz}^{(1)}, g_{\sss hzz}^{(2)}, g_{\sss haz}^{(1)}, g_{\sss haz}^{(2)},
       g_{\sss hww}^{(1)}, g_{\sss hww}^{(2)} \Big\}$ \\
    $g^{(3)}_{\sss hhzz}$  &$\frac{g^2}{2 \cW^2} \Big[ 1 - 6 \bar c_{\sss T} - \bar c_{\sss H} + 8 \bar c_{\sss\gamma} \frac{\sW^4}{\cW^2}\Big]$ \\
    $g^{(1)}_{\sss haww}$ & $\frac{g^2 \sW}{\mW} \Big[2 \bar c_{\sss W} + \bar c_{\sss HB} +\bar c_{\sss HW} \Big]$ \\
    $\tilde g^{(1)}_{\sss haww}$ & $\frac{g^2 \sW}{\mW} \Big[\tilde c_{\sss HW} -\tilde c_{\sss HB} \Big]$ \\
    $g^{(2)}_{\sss haww}$ & $\frac{2 g^2 \sW}{\mW} \bar c_{\sss W}$ \\
    $\tilde g^{(2)}_{\sss haww}$ & $\frac{g^2 \sW}{\mW} \tilde c_{\sss HW}$ \\
    $g^{(3)}_{\sss haww}$ & $\frac{g^2 \sW}{\mW} \Big[ \bar c_{\sss HW} + \bar c_{\sss W}\Big]$ \\
    $g^{(1)}_{\sss hzww}$ & $\frac{g^2}{\cW \mW} \Big[\bar c_{\sss HW} \cW^2 - \bar c_{\sss HB} \sW^2 +
       \bar c_{\sss W} (3-2\sW^2) \Big]$ \\
    $\tilde g^{(1)}_{\sss hzww}$ & $\frac{g^2}{\cW \mW} \Big[ \tilde c_{\sss HW} (2-\sW^2) + \tilde c_{\sss HB} \sW^2\Big]$ \\
    $g^{(2)}_{\sss hzww}$ & $\frac{g^2}{\cW \mW} \Big[\bar c_{\sss HW} + \bar c_{\sss W}(3-2 \sW^2) \Big]$ \\
    $\tilde g^{(2)}_{\sss hzww}$ & $\frac{2 g^2}{\mW} c_{\sss W} \tilde c_{\sss HW} $ \\
    $g^{(3)}_{\sss hzww}$ & $\frac{g^2}{\cW \mW} \sW^2 \Big[ \bar c_{\sss HW} + \bar c_W\Big]$ \\
  \end{tabular}
  \caption{Quartic interactions of one or several Higgs field with gauge bosons.
    We present the relations between the Lagrangian parameters introduced in Eq.~\eqref{eq:massbasis4}, where
    the Lagrangian is expressed in the mass basis, and those associated with the operators
    of Section~\ref{sec:HELgauge} expressed in the gauge basis.}
  \label{tab:params4hhvv}
\end{table}
\renewcommand{\arraystretch}{1}

Both Lagrangians ${\cal L}_3$ and ${\cal L}_4$
not only exhibit Lorentz
structures common with the Standard Model interactions, but also some novel ones. For instance, focusing on
the Higgs to $W$-boson trilinear interactions, the complete Feynman rule reads
\begin{center}\begin{tabular}{r l}
  \parbox{0.25\textwidth}{
    \vspace*{-1cm}
    \includegraphics[width=.25\columnwidth]{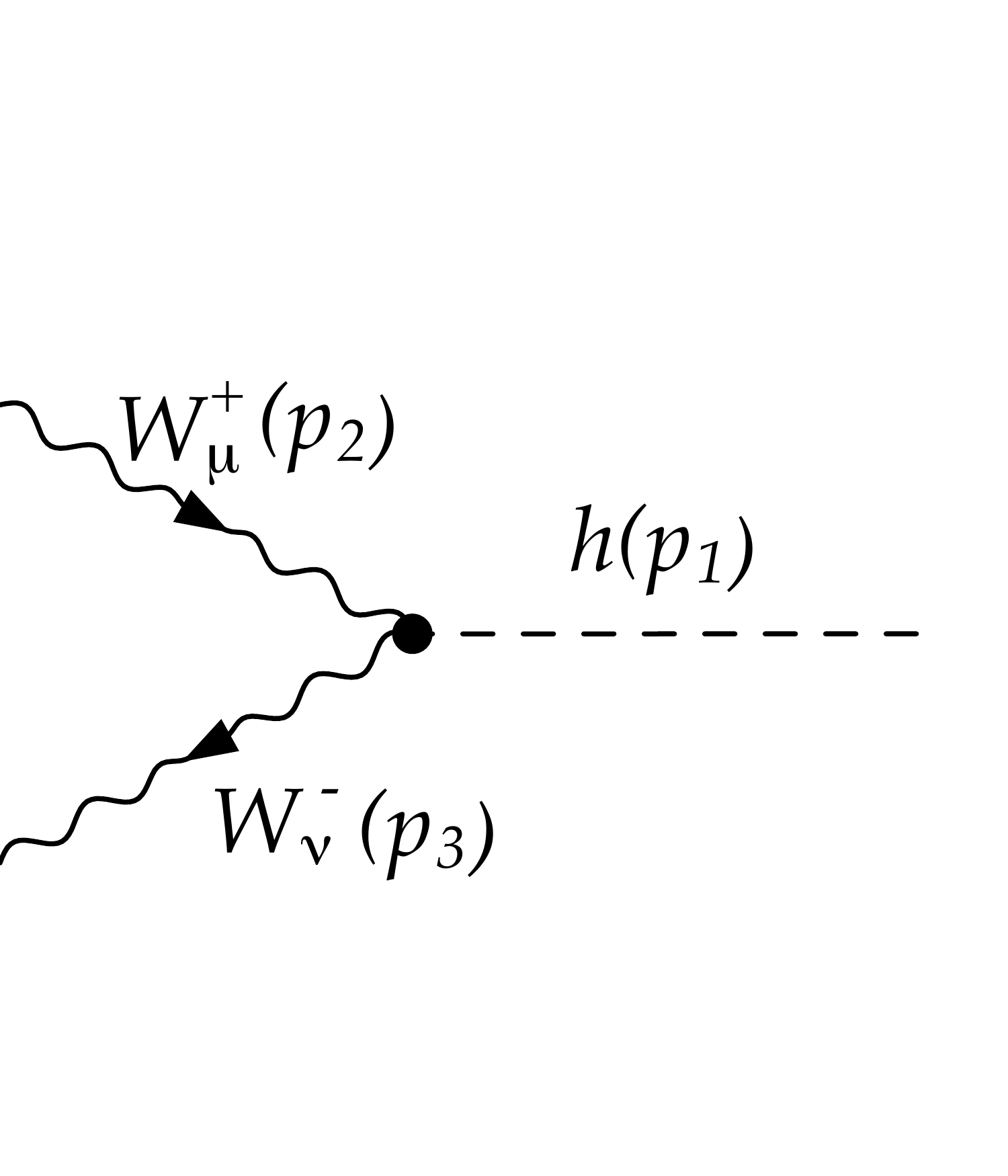}
    \vspace*{-1cm}
    }
  &
\parbox{0.7\textwidth}{$\displaystyle 
   i \Big[
       \eta^{\mu\nu} \big(g \mW + g_{\sss hww}^{(1)} p_2\cdot p_3 + g_{\sss hww}^{(2)} (p_2^2+p_3^2)\big) \\
       ~~~ -g_{\sss hww}^{(1)} p_2^\nu p_3^\mu
       -g_{\sss hww}^{(2)} \big(p_2^\nu p_2^\mu+p_3^\nu p_3^\mu\big)
       - \epsilon^{\mu\nu\rho\sigma} \tilde g_{\sss hww} p_{2\rho} p_{3\sigma}
   \Big] $\ ,}
\end{tabular}\end{center}
where only the component proportional to the metric is present in the Standard Model.

Finally, new higher-dimensional operators involving the Higgs field can also have implications on vertices
describing the self-interactions of the
gauge bosons, in particular once the neutral component of the $SU(2)_L$ doublet $\Phi$ gets
its vacuum expectation values. These new contributions supplement those
accounted for by the Lagrangians ${\cal L}_{CP}$ and ${\cal L}_G$ of Eq.~\eqref{eq:silhCPodd} and
Eq.~\eqref{eq:lagG} that we omit from the rest of the
discussion of this
subsection for brevity. The complete and lengthy expressions can however be obtained from the {\sc FeynRules}
implementation of the model.
The new physics effects that are induced by all the other operators of the Lagrangian of
Eq.~\eqref{eq:effL} are then given, including the Standard Model contributions, by
\be\bsp
  {\cal L}_{3V} = &\ \Big[ i g_{\sss aww}^{(1)} W^\dag_{\mu\nu} A^\mu W^\nu \hc \Big]
   + i g_{\sss aww}^{(2)} F_{\mu\nu} W^\mu W^{\nu\dag}
\\&\
  + \Big[ i g_{\sss zww}^{(1)} W^\dag_{\mu\nu} Z^\mu W^\nu \hc \Big]
   + i g_{\sss zww}^{(2)} Z_{\mu\nu} W^\mu W^{\nu\dag}
 \ ,
\esp\label{eq:l3v}\ee
and
\be\bsp
  {\cal L}_{4V} =&\
     g_{\sss wwww} \Big[ W_\mu W^\mu W^\dag_\nu W^{\nu\dag} - W_\mu W^\nu W^\dag_\nu W^{\mu\dag} \Big]
  \\ &\
   + g_{\sss aaww} \Big[ A^\mu A_\mu W^\dag_\nu W^\nu - A_\mu A^\nu W^\dag_\nu W^\mu \Big]
  \\&\
   + g_{\sss zzww} \Big[ Z^\mu Z_\mu W^\dag_\nu W^\nu - Z_\mu Z^\nu W^\dag_\nu W^\mu \Big]
  \\&\
   + g_{\sss azww} \Big[ A^\mu Z_\mu W^\dag_\nu W^\nu - A_\mu Z^\nu W^\dag_\nu W^\mu \Big]
  \ , 
\esp\label{eq:l4v}\ee
after splitting the different terms into three-point and four-point contributions, respectively, and neglecting
any other vertex containing more than four external legs.
We collect the coefficients of the different operators entering those Lagrangians in Table~\ref{tab:paramsL3v} and
Table~\ref{tab:paramsL4v} and give their expressions in terms of the Wilson coefficients of the Higgs effective Lagrangian
presented in Section~\ref{sec:HELgauge}.

\renewcommand{\arraystretch}{1.5}
\begin{table}
  \center
  \begin{tabular}{l| l }
  \hline\hline
    Eq.~\eqref{eq:l3v} & 
    Section~\ref{sec:HELgauge}\\
  \hline \hline
    $g_{\sss aww}^{(1)}$ & $e \Big[ 1 -  2 \bar c_{\sss W} \Big]$ \\
    $g_{\sss aww}^{(2)}$ & $e \Big[ 1 -  2 \bar c_{\sss W} - \bar c_{\sss HB} - \bar c_{\sss HW} \Big]$ \\
    $g_{\sss zww}^{(1)}$ & $\frac{g}{\cW} \Big[ \cW^2 -  \bar c_{\sss HW} + (2\sW^2-3) \bar c_{\sss W} \Big]$ \\
    $g_{\sss zww}^{(2)}$ & $\frac{g}{\cW} \Big[ \cW^2(1-\bar c_{\sss HW}) + \sW^2 \bar c_{\sss HB}  +
(2\sW^2-3) \bar c_{\sss W} \Big]$ \\
  \end{tabular}
  \caption{Trilinear gauge interactions.
    We present the relations between the Lagrangian parameters introduced in Eq.~\eqref{eq:l3v}, where
    the Lagrangian is expressed in the mass basis, and those associated with the operators
    of Section~\ref{sec:HELgauge} expressed in the gauge basis.}
  \label{tab:paramsL3v}
\end{table}
\renewcommand{\arraystretch}{1}

\renewcommand{\arraystretch}{1.5}
\begin{table}
  \center
  \begin{tabular}{l| l }
  \hline\hline
    Eq.~\eqref{eq:l4v} & 
    Section~\ref{sec:HELgauge}\\
  \hline \hline
    $g_{\sss wwww}$ & $\frac{g^2}{2} \Big[ 1 - 2 \bar c_{\sss HW} - 4 \bar c_{\sss W} \Big]$ \\
    $g_{\sss aaww}$ & $e^2 \Big[ -1 +  2 \bar c_{\sss W} \Big]$ \\
    $g_{\sss zzww}$ & $g^2 \Big[ -\cW^2 +  2 \bar c_{\sss HW} + 2 (2-\sW^2) \bar c_{\sss W}  \Big]$ \\
    $g_{\sss azww}$ & $\frac{2ge}{\cW} \Big[ -\cW^2 + \bar c_{\sss HW} + (3-2\sW^2) \bar c_{\sss W}  \Big]$ \\
  \end{tabular}
  \caption{Quartic gauge interactions.
    We present the relations between the Lagrangian parameters introduced in Eq.~\eqref{eq:l4v}, where
    the Lagrangian is expressed in the mass basis, and those associated with the operators
    of Section~\ref{sec:HELgauge} expressed in the gauge basis.}
  \label{tab:paramsL4v}
\end{table}
\renewcommand{\arraystretch}{1}

\section{Experimental constraints on dimension-six effective operators}\label{sec:bounds}

The magnitude of the Wilson coefficients associated with the dimension-six operators
introduced in Section~\ref{sec:HEL}, and thus an estimate of their impact
on physical observables, can be naively computed by a simple power
counting~\cite{Giudice:2007fh,Contino:2013kra}. In this way, each power
of $\Phi$ leads to a $g_{NP}/M$ suppression factor, $M$ being the typical
mass scale of the new physics sector and $g_{NP}$ the coupling strength of the new
states to the Higgs field, while each derivative implies an additional reduction
of $1/M$. Additionally, in the framework of a given theory, specific operators can be generated
at the one-loop level so that additional suppression can be foreseen.
Sticking to tree-level, Wilson coefficients such as $\bar c_{\sss H}$, $\bar c_{\sss T}$, $\bar c_{\sss 6}$
or $\bar c_\psi$ are expected to be of the order of
\be
  \bar c_{\sss H}, \bar c_{\sss T}, \bar c_{\sss 6}, \bar c_\psi
    \sim {\cal O}\left( \frac{g_{NP}^2 v^2}{M^2} \right) \ ,
\ee
and can therefore be quite large for strongly coupled new physics,
while in contrast, the coefficients of operators such as $\bar c_{\sss W}$ and $\bar c_{\sss B}$
scale as
\be
  \bar c_{\sss B}, \bar c_{\sss W}
    \sim {\cal O}\left( \frac{\mW^2}{M^2} \right) \ ,
\ee
and are thus expected to be relatively suppressed or enhanced according to the value
of the ratio $g/g_{NP}$.

In addition, the value of the 39 Wilson coefficients introduced in
Section~\ref{sec:HEL} is experimentally constrained from various sources.
These limits mainly arise from collider data (LEP, Tevatron, LHC), electric dipole moment
measurements, rare decay bounds and results of several experiments dedicated to the measurement
of the anomalous
magnetic moments of the muon and the electron.
An exhaustive list of the current bounds on the various coefficients of the
considered dimension-six operators can be found in Ref.~\cite{Contino:2013kra}
so that we omit it from the present manuscript.

Some of these constraints however only
involve specific combinations of Wilson coefficients instead of a single coefficient.
An example lies in the recasting, in the language
of the Higgs effective Lagrangian of Section~\ref{sec:HELgauge}, of the constraints originating from
the electroweak precision parameters derived in Ref.~\cite{Baak:2012kk},
\be
  \bar c_{\sss T}(m_Z) \in [-1.5,2.2] \times 10^{-3} \qquad\text{and}\qquad
  \Big(\bar c_{\sss W}(m_Z) + \bar c_{\sss B}(m_Z)\Big) \in [-1.4,1.9] \times 10^{-3} \ .
\ee
This still leaves open the possibility of a cancellation between
two large $\bar c_W$ and $\bar c_B$ coefficients at the $Z$-pole. The possibility of such
cancellations also holds for other coefficients, such as $\bar c_{\sss HW}$ or $\bar c_{\sss HB}$,
for which there exists no strong bound from the LEP experiments. A more involved example
arises from the strong limits on the quantity $\bar c_{\sss WW}$
defined by
a combination of several of the  $\bar c_{\sss i}$ coefficients,
\be
  {\bar c}_{\sss WW} = {\bar c}_{\sss W} - {\bar c}_{\sss B} + {\bar c}_{\sss HB}
     - {\bar c}_{\sss HW} + \frac14 {\bar c}_{\sss \gamma} \ ,
\ee
Existing data being too limited to allow for
disentangling the individual effects of the different operators, this implies that no constraints can be really
inferred in our choice of basis for the dimension-six operators.

Those considerations motivate us to avoid a careful design of a benchmark scenario
theoretically motivated and not experimentally excluded, a task that can only be performed
from a global fit of all data and considering all dimension-six operators,
including additional four-fermion interactions, their one-loop
mixings as well as field equations of motion linking our basis of operators
to an extended one with redundant operators. We instead pick up, in Section~\ref{sec:pheno},
 several phenomenological
examples illustrating possible usages of the Higgs effective field theory implementation
achieved in this work, so that it could be possibly used for such global fits in future works.

\section{Phenomenological examples}
\label{sec:pheno}

\subsection{Interactions between the Higgs boson and the electroweak gauge bosons}\label{sec:HtoVV}

As first examples of a possible usage of our implementation, we focus
on the restricted set of operators implying modifications of the $hVV$ couplings
between one single
Higgs boson and a gauge-boson pair. New physics effects
possibly arising in such interactions have
been largely investigated in pioneering and more recent works prior
the discovery of a Higgs boson by the ATLAS and CMS experiments~\cite{Dell'Aquila:1985ve,%
Dell'Aquila:1985vc,Dell'Aquila:1985vb,Dell'Aquila:1988rx,%
Dell'Aquila:1988fe,DeRujula:1991se,%
Barger:1990mn,Hagiwara:1992eh,Hagiwara:1993ck,Dittmar:1996ss,Dittmar:1997nea,GonzalezGarcia:1999fq,Eboli:1999pt,%
Choi:2002jk,Odagiri:2002nd,Buszello:2002uu,Djouadi:2005gi,Buszello:2006hf,%
Bredenstein:2006rh,BhupalDev:2007is,Godbole:2007cn,Hagiwara:2009wt,DeRujula:2010ys,%
Englert:2010ud,DeSanctis:2011yc,Barger:2011cb,Bonnet:2011yx,%
Ellis:2012wg,Biswal:2012mp}. Within the last two years,
they have been reassessed in the light of
newer LHC data including the Higgs results~\cite{Corbett:2012dm,Bolognesi:2012mm,Boughezal:2012tz,%
Stolarski:2012ps,Ellis:2012jv,Choi:2012yg,Masso:2012eq,%
Corbett:2012ja,Modak:2013sb,Corbett:2013pja,Dumont:2013wma,Frank:2013gca,Corbett:2013hia}.
We reproduce and extend, in the rest of this subsection, some of these results by employing the
{\sc MadGraph}~5 package~\cite{Alwall:2011uj},
using our {\sc FeynRules}~\cite{Christensen:2008py,
Alloul:2013bka} implementation to generate the necessary
UFO model files~\cite{Degrande:2011ua}. The events have eventually been
analyzed by means of {\sc MadAnalysis}~5~\cite{Conte:2012fm,Conte:2013mea}.

\subsubsection{Probing the custodial symmetry}
Among all the Wilson coefficients included in the Lagrangian of Eq.~\eqref{eq:effL},
the $\bar c_{\sss W}$ parameter has the specificity to imply modifications
of both the $hZZ$ and $hWW$ interactions simultaneously.
Its value can therefore be probed by independent
measurements of the Higgs boson properties in its $WW$ and $ZZ$ decay modes,
when considering, \eg, leptonic weak boson decays,
\be\label{eq:dechiggs}
  h \to Z^* Z^{(*)} \to 4\, \ell
  \qquad \text{and}\qquad
  h \to W^* W^{(*)} \to  \ell \nu \ell \nu \ .
\ee
Analyzing the full set of LHC data describing collisions at center-of-mass energies
of 7~TeV and 8~TeV, considering those two decay patterns and a Higgs-boson production
by gluon fusion,
the ATLAS collaboration has recently\footnote{A similar study has been performed by the
CMS collaboration, but the results do not account for all 2011-2012 LHC
data~\cite{CMS:yva}.} reported a measurement~\cite{Aad:2013wqa}
of the ratio of the two corresponding branching fractions estimated relatively to
the Standard Model expectations,
\be
  \lambda_{\sss WZ} = \frac{\kappa_{\sss W}}{\kappa_{\sss Z}}  \ ,
\ee
with
\be
  \kappa_{\sss W} = \frac{\Gamma(h \to W^* W^{(*)})}{\Gamma(h \to W^* W^{(*)})_{SM}}
  \qquad \text{and}\qquad
  \kappa_{\sss Z} = \frac{\Gamma(h \to Z^* Z^{(*)})}{\Gamma(h \to Z^* Z^{(*)})_{SM}} \ .
\ee
Assuming a Higgs boson mass of $m_h = 125.5$~GeV, the quantity $\lambda_{\sss WZ}$ has been found to lie
in a range defined by
\be
  \lambda_{\sss WZ} = 0.82 \pm 0.15\ ,
\ee
at the 95\% confidence level and when keeping the strength of the Higgs boson to two photon
coupling as a free parameter. This last assumption allows for beyond the Standard Model
effects in the $h\gamma\gamma$ interactions, the latter being in particular
expected from a non-zero $\bar c_{\sss W}$
parameter too (see Table~\ref{tab:paramshvv}).
Employing our framework, we theoretically estimate the two $\kappa$ parameters and the
$\lambda_{\sss WZ}$ quantity as
\be\bsp
  \kappa_{\sss W} = &\ 1+ 2.23 \bar c_{\sss W} + 1.27 \, \bar c_{\sss W}^2 \ , \\
  \kappa_{\sss Z} = &\ 1+ 1.97 \, \bar c_{\sss W}  + 1.00 \, \bar c_{\sss W}^2\ , \\
  \lambda_{\sss WZ}=&\ 1+ 0.28 \, \bar c_{\sss W}  - 0.27 \, \bar c_{\sss W}^2 \ ,
\esp \label{eq:kaplam}\ee
so that we extract, at the 95\% confidence level,
\be
 \bar c_{\sss W} \in [-1.71, 0.42] \ ,
\ee
from the ATLAS results.
This range is still largely compatible with the naive estimate computed as indicated
in Section~\ref{sec:bounds},
\be
  \bar c_{\sss W} \simeq {\cal O}\left( \frac{v}{\Lambda}\right)^2 \simeq10^{-3} \ ,
\ee
considering TeV scale new physics~\cite{Corbett:2012dm,Masso:2012eq,%
Corbett:2012ja,Corbett:2013pja,Dumont:2013wma,Corbett:2013hia,Contino:2013kra} and also agrees
with predictions performed by making use of the {\sc eHDecay} program~\cite{Contino:2013kra},
the extension of the {\sc HDecay} package~\cite{Djouadi:1997yw,Butterworth:2010ym}
in the framework of the Higgs effective field theory of Section~\ref{sec:HELgauge}.

In the case both the $\bar c_{\sss W}$ and $\bar c_{\sss B}$ coefficients are non-vanishing,
as suggested by the LEP limits constraining their difference, the additional effect of the
$\bar c_{\sss B}$ parameter on the $\kappa_{\sss Z}$ quantity can be obtained
by replacing $\bar c_{\sss W}$ by
$\bar c_{\sss W} +\tan^2 \theta_W \bar c_{\sss B}$ in
Eq.~\eqref{eq:kaplam}. The corresponding modification in the predictions
for the $\lambda_{\sss WZ}$ parameter is trivially obtained by, \eg, linearizing
the new physics effects.

\begin{figure}[t]
\centerline{\includegraphics[height=7cm]{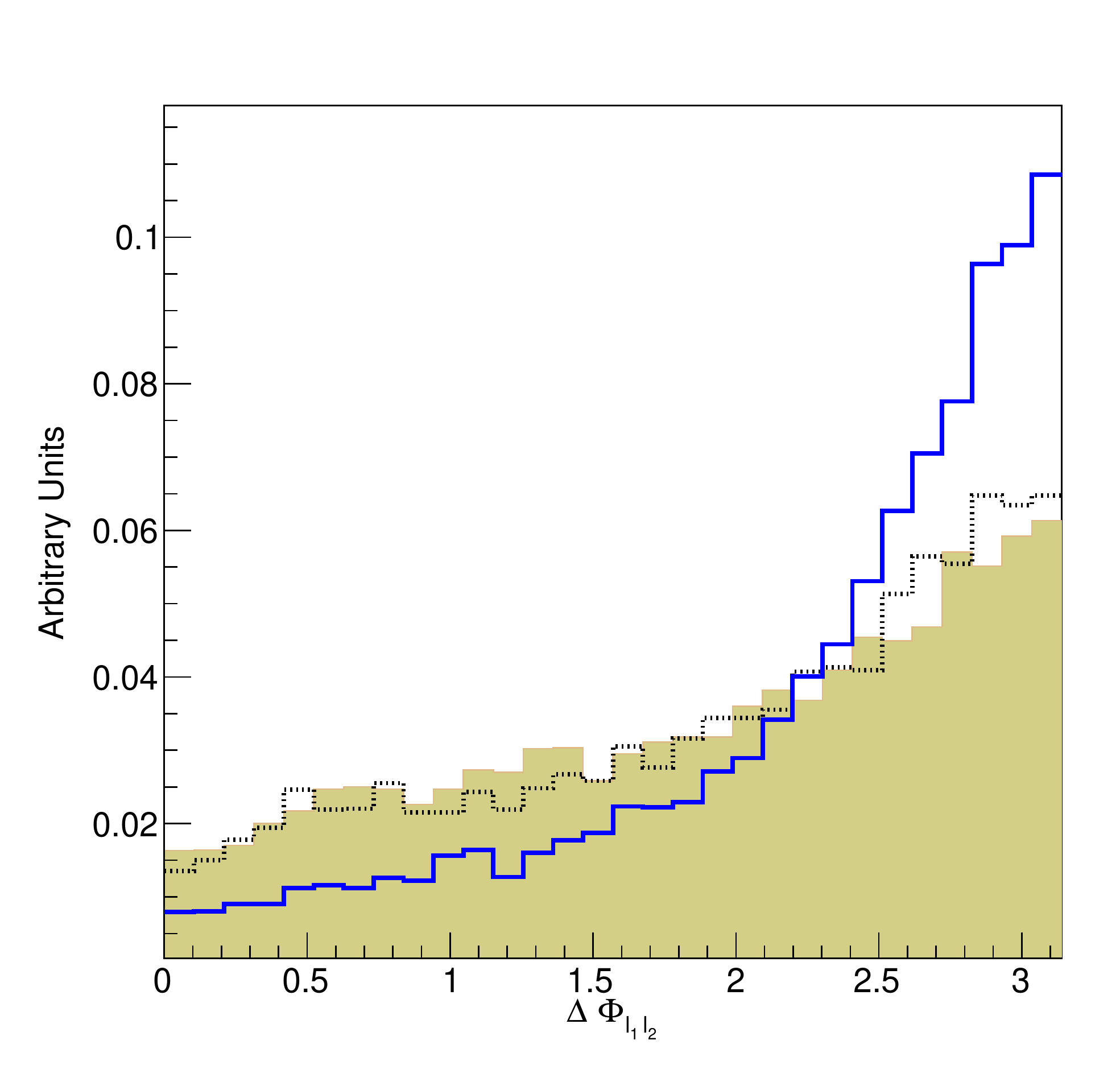}}
\caption{
{\it Distribution of the angular separation of the decay planes associated with each of the (possibly
virtual) $Z$-bosons resulting from the decay of a Higgs boson produced in the gluon fusion mode.
We show the Standard model distribution (green-solid histogram) to which we superimpose
predictions associated with a non-zero $\bar{c}_{\sss W}$= 0.3 (black-dotted line)
and -1.5 (blue-solid line) Wilson coefficient.}}
\label{fig:delphi}
\end{figure}

\subsubsection{Kinematics of the four-lepton system issued from a $h \to Z^* Z^{(*)}$ decay}
In addition to shifting the values of the Higgs signal strengths $\kappa_{\sss W}$ and
$\kappa_{\sss Z}$, higher-dimensional operators also affect various kinematical distributions.
In particular, several angular distributions related to the
leptonic systems arising from decays
of the Higgs boson into four fermions such as those introduced in Eq.~\eqref{eq:dechiggs}
can be drastically modified by new physics effects.
Considering the $ZZ^*$ channel in the four-lepton mode
where the entire final state can be reconstructed, we simulate LHC collisions
at a center-of-mass energy of 14~TeV yielding the production of a Higgs boson
from a gluon-pair initial state, followed by its decay into four leptons,
\be
  g \, g \to h \to Z^* Z^{(*)} \to (\ell_1^- \ell^+_1) \, (\ell^-_2 \ell^+_2) \ .
\ee
In our setup, each lepton is required to have a transverse momentum greater than 10~GeV,
a pseudorapidity $|\eta| < 2.5$ and to be isolated.
We define lepton isolation by enforcing a minimum angular distance $\Delta R$ of 0.4 between
any two of the produced particles. In our notations,
$\Delta R = \sqrt{\Delta\varphi^2 + \Delta\eta^2}$ with $\varphi$ denoting the azimuthal angle with
respect to the beam direction.

The four-momenta of the four final state leptons can be seen as lying on three distinct planes,
one of them being related to the decaying Higgs boson into two (possibly virtual)  $Z$-bosons
and the two other being associated with the decay products of each $Z$-boson. As it is customary in the
literature~\cite{Dell'Aquila:1985ve,Dell'Aquila:1985vc,Dell'Aquila:1985vb,Dell'Aquila:1988rx,%
Dell'Aquila:1988fe,Choi:2002jk,Odagiri:2002nd,Buszello:2002uu,Djouadi:2005gi,Buszello:2006hf,%
Bredenstein:2006rh,BhupalDev:2007is,Godbole:2007cn,Hagiwara:2009wt,DeRujula:2010ys,%
Englert:2010ud,DeSanctis:2011yc,Barger:2011cb,Bolognesi:2012mm,Boughezal:2012tz,%
Stolarski:2012ps,Choi:2012yg,Modak:2013sb}, we denote by $\theta_{1,2}$ the polar angles
of the two leptons $\ell^-_{1,2}$ in the rest frame of the respective decaying $Z$-bosons,
and by $\Delta\phi_{\ell_1^- \ell^-_2}$ the azimuthal angle between the planes formed by
each lepton pair in the Higgs boson rest frame.

We show in Figure~\ref{fig:delphi} the angular distribution $\Delta \phi_{\ell_1 \ell_2}$
for the pure Standard model case (green-solid histogram) to which we superimpose predictions obtained
for two choices of non-vanishing $\bar{c}_{\sss W}$ Wilson coefficient. These latter predictions
also include Standard Model contributions with which new physics interferes.
The black-dashed and blue-solid lines in
the figure correspond to values of $\bar{c}_{\sss W}$= 0.3 and -1.5, respectively.
The Standard Model spectrum is expected to be a function of both $\cos(\Delta \phi_{\ell_1 \ell_2})$
and $\cos(2 \Delta \phi_{\ell_1 \ell_2})$ exhibiting a rather gentle slope. This known
result~\cite{Barger:1993wt,Godbole:2007cn} is
recovered by our predictions. In contrast, the structure of the ${\cal O}_{\sss W}$
dimension-six operators modifies the
coefficients of the polynomial in the cosines of the $\Delta\phi_{\ell_1 \ell_2}$ angle, leading to
a steeper dependence. The larger the value of $\bar{c}_{\sss W}$, the steeper this function becomes.
Consequently, this $\Delta\phi_{\ell_1 \ell_2}$ distribution can clearly be used,
if measured with a decent precision,
to constrain the magnitude of the Wilson coefficient associated with the operator ${\cal O}_{\sss W}$.

\subsubsection{Kinematics of the two-lepton system issued from a $h \to W^* W^{(*)}$ decay}

\begin{figure}[t]
  \includegraphics[width=.49\columnwidth]{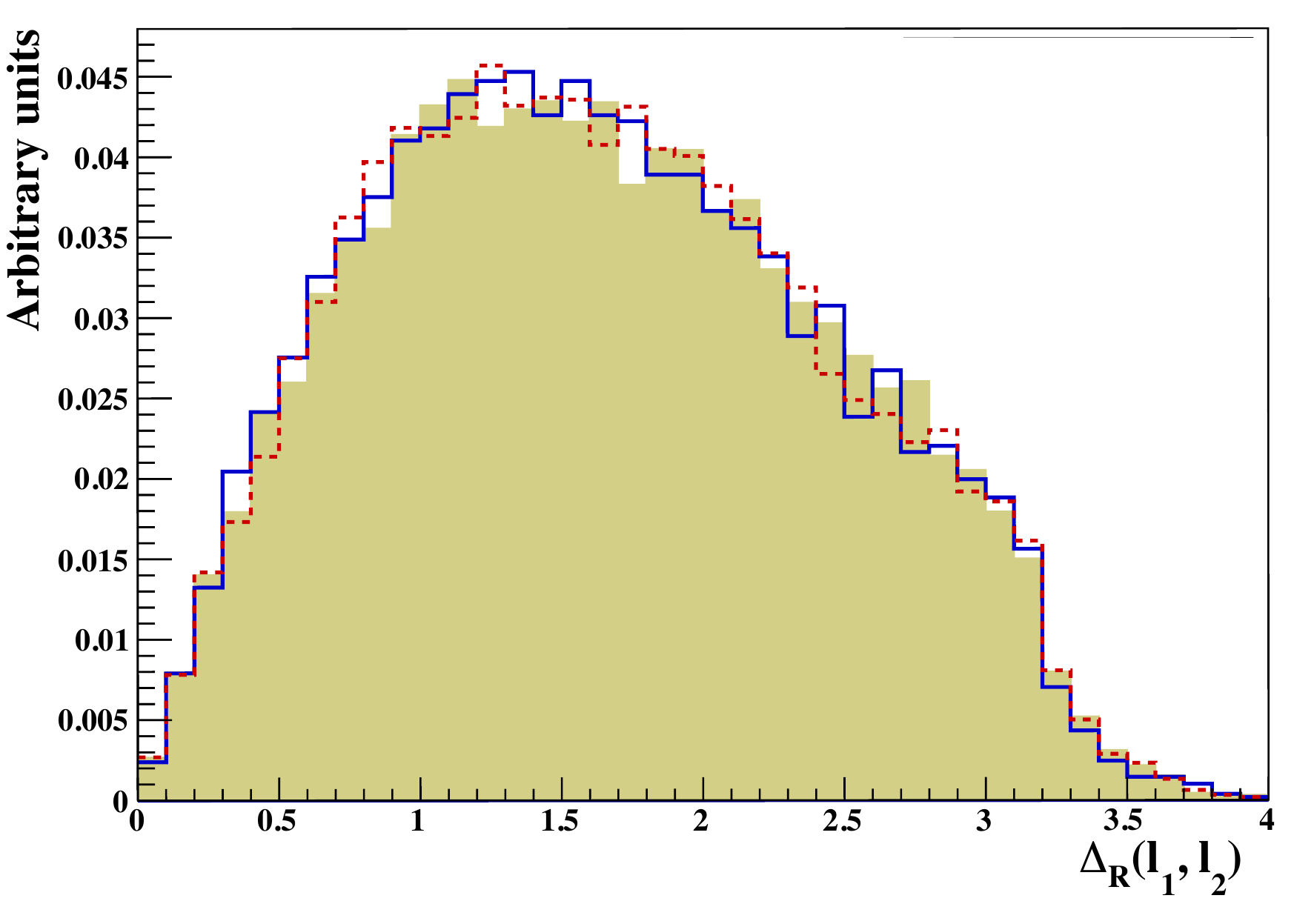}
  \includegraphics[width=.49\columnwidth]{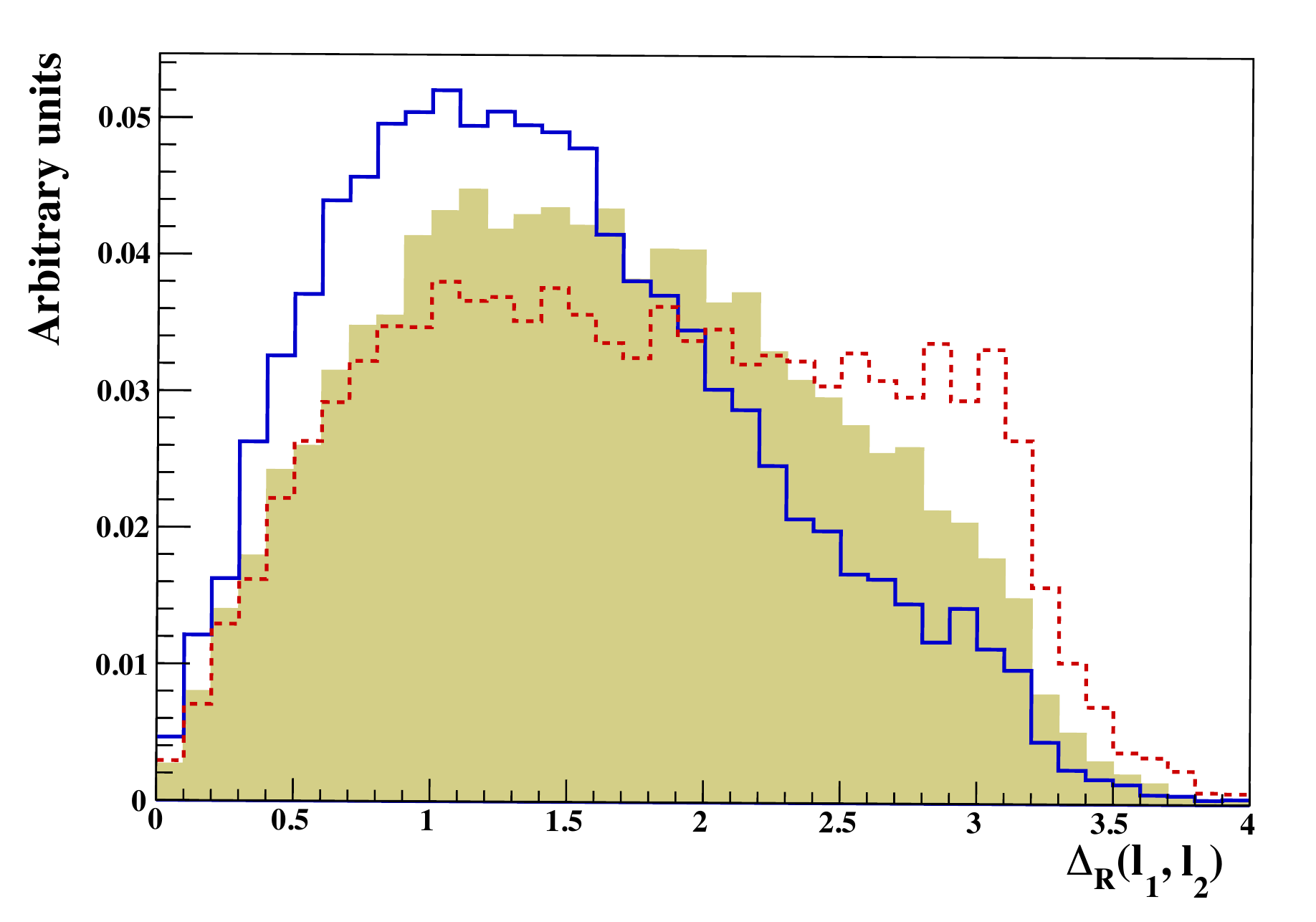}
    \caption{{\it Distribution of the angular distance $\Delta R(\ell_1,\ell_2)$
    between the two charged leptons originating from a pair of $W$-bosons issued from the decay
    of a Higgs boson produced via gluon fusion. The left panel of the figure illustrates that for
    $\bar{c}_W = 0.05$ (red-dashed line) and $\bar{c}_{HW} = 0.05$ (blue-solid line),
    no sizable effect beyond the SM predictions (green-solid histogram) can be expected.
    In contrast, the right panel of the figure
    shows that for larger values of the Wilson coefficients set to $-1$, sensitive effects can be observed.}}
\label{fig:hww}
\end{figure}

We now consider the $WW^*$ Higgs decay mode in the case both weak bosons decay leptonically,
\be
  g\, g\to h\to W^* W^{(*)} \to 2 \, \ell \, 2 \, \nu \ ,
\ee
in the framework of LHC collisions at a center-of-mass energy of 14~TeV and for similar lepton
requirements as in the previous subsection.
We study new physics effects arising from non-vanishing $\bar{c}_{\sss W}$ and $\bar{c}_{\sss HW}$
parameters in the spectrum of the
angular distance $\Delta R$, in the $(\eta,\varphi)$ plane, among the two produced leptons.
We have found that the resulting distribution is barely affected when the two Wilson
coefficients of interest lie in the range $[-0.5, 0.5]$. This feature is illustrated on the left panel
of Figure~\ref{fig:hww} in which we first compute the Standard Model expectation obtained
when all the new physics parameters are set to zero (green-solid histogram)
and then superimpose results obtained for the choices of $\bar{c}_{\sss HW} = 0.5$ (blue-solid line)
and $\bar{c}_{\sss W} = 0.05$ (red-dashed line). In contrast, for larger
values of the Wilson coefficients, important differences in the distributions
can be expected, as shown on the right panel of the figure for two
representative scenarios with respectively $\bar{c}_{\sss HW} = -1$ (blue-solid line)
and $\bar{c}_{\sss W} = -1$ (red-dashed line). It can be noted that
according to the new physics benchmark scenarios
the most striking effects can be either expected in the low $\Delta R$ region, or in the higher one.

\begin{figure}[t]
\centering
 \includegraphics[width=.49\columnwidth]{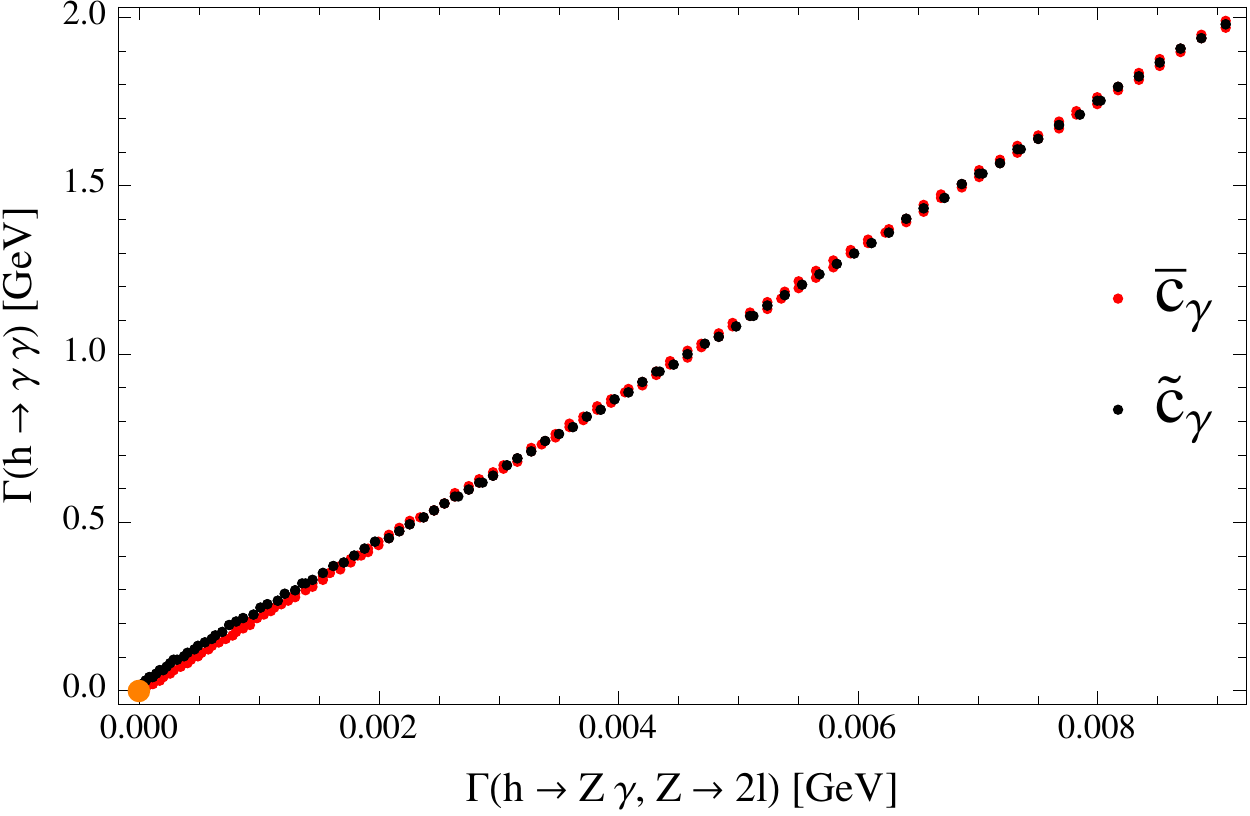}
 \includegraphics[width=.49\columnwidth]{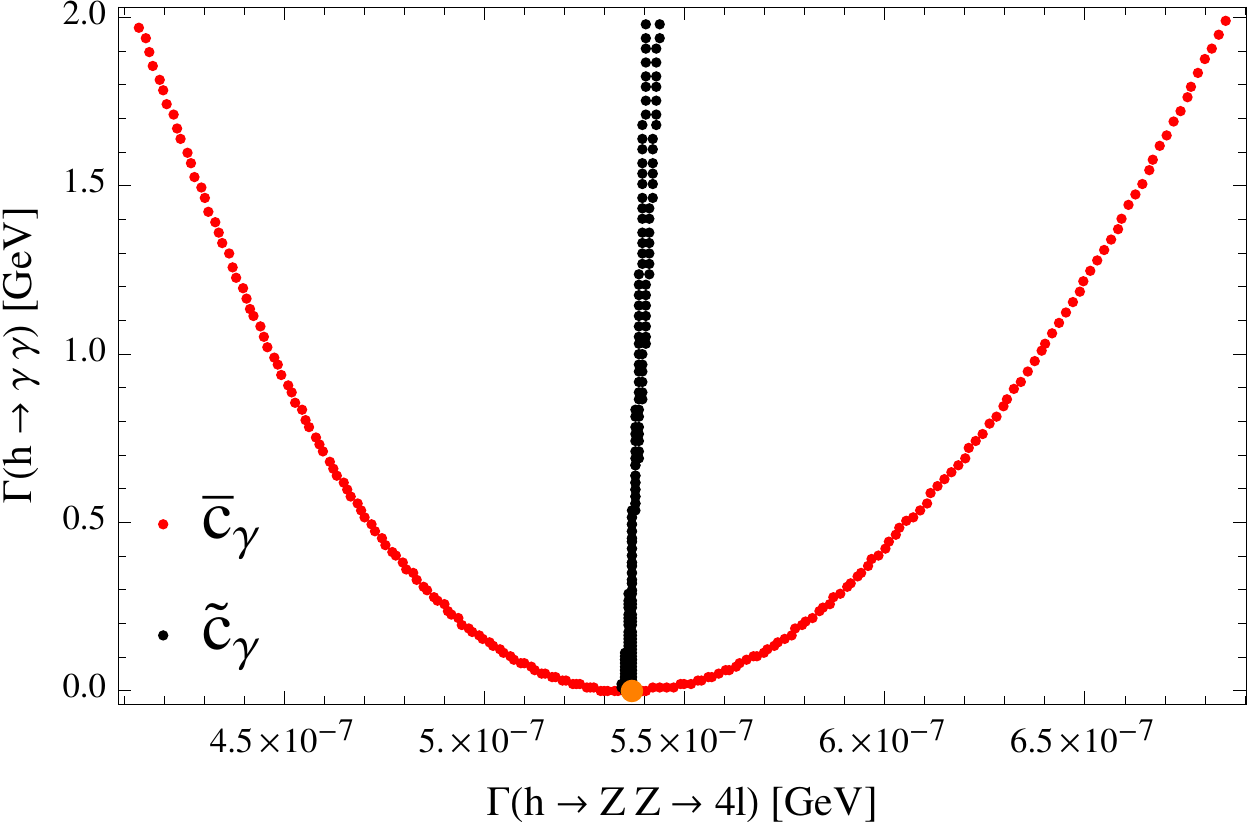}\\ \vspace{.5cm}
 \includegraphics[width=.49\columnwidth]{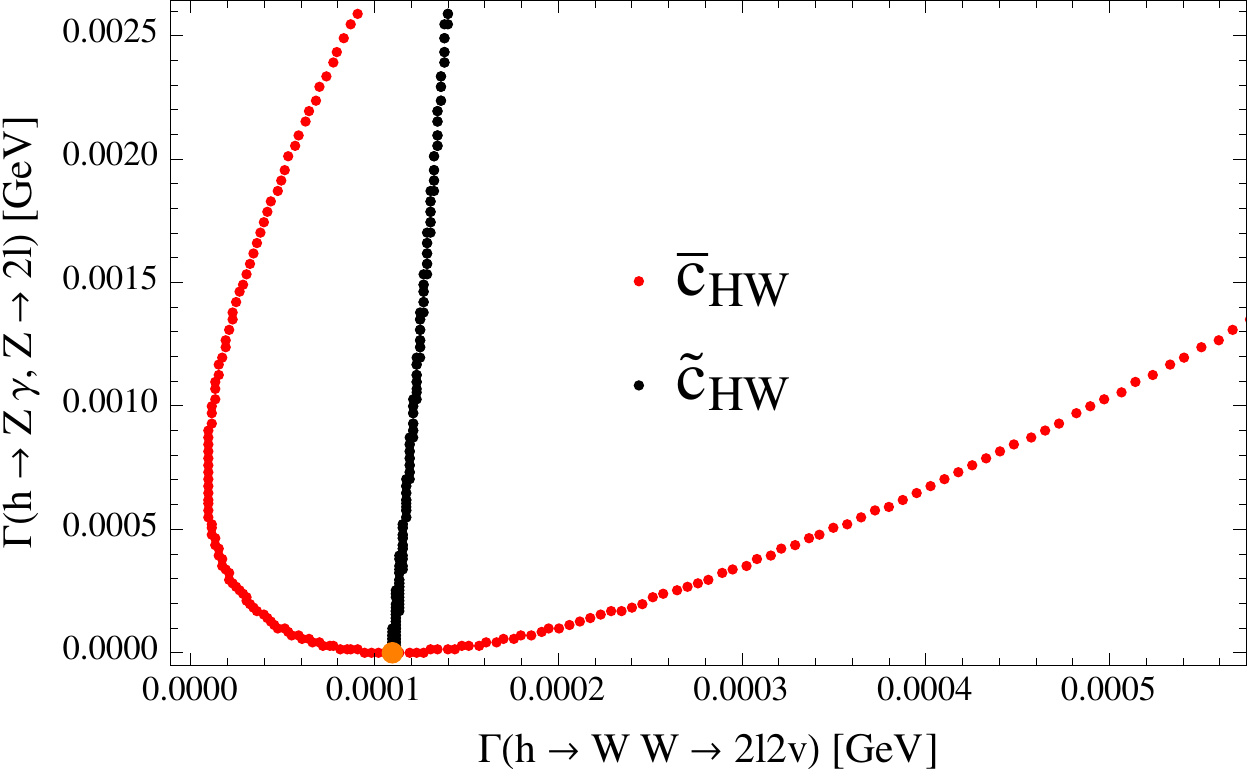}
 \includegraphics[width=.49\columnwidth]{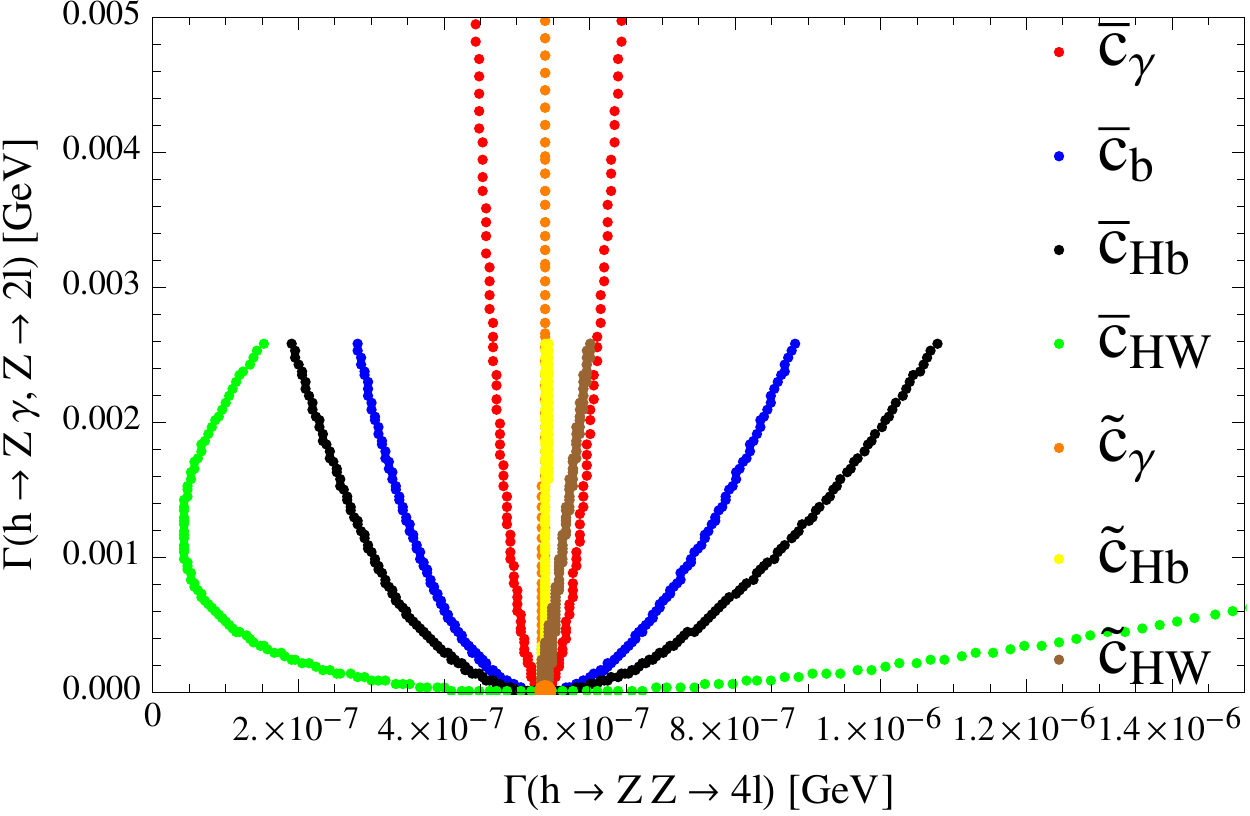}\\ \vspace{.5cm}
 \includegraphics[width=.49\columnwidth]{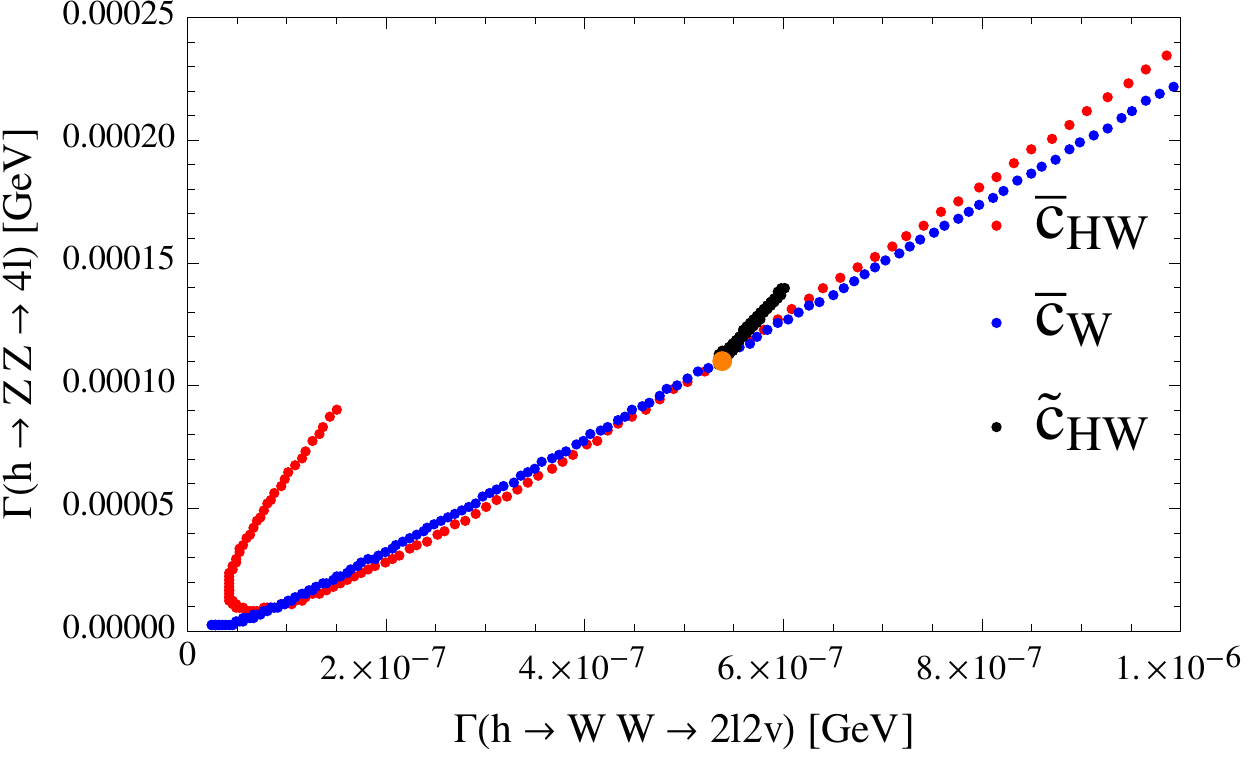}
\caption{\textit{Correlations of several Higgs boson partial widths when different Wilson coefficients
  are chosen non-vanishing. In the upper-left (upper-right) panel of the figure, we investigate simultaneous
  new physics effects on the $h\to \gamma \, \gamma$ and $h\to Z\, \gamma$ ($h\to Z\, Z$) partial widths, in its
  central-left (central-right) panel on the $h\to Z\, \gamma$ and
  $h\to W \, W$ ($h\to Z \, Z$) partial widths and in its lower panel on the $h\to Z \, Z$ and $h\to W\, W$ partial
  widths. We indicate the Standard Model expectation by an orange dot.}}
\label{fig:correlations}
\end{figure}

\subsubsection{Correlations of new physics effects in the Higgs boson partial widths}

As already mentioned in Section~\ref{sec:HEL}, one given dimension-six operator may affect several interactions
of the Higgs boson to a pair of gauge bosons. This feature is employed in
Figure~\ref{fig:correlations} where we compute different Higgs boson partial widths when considering
a specific dimension-six operator, allowed to vary in the $[-1,1]$ range.
The results of the different decay channels are then confronted to each other
and we study the correlations among the corresponding partial widths. It can be seen
that this information could be used in the future to constrain
which Wilson coefficients are allowed to be non-negligible by data.
It is however clear that this type of investigation
consists only of a first step towards a global fit of a complete set of
dimension-six operators. In addition, our results are compared in each of the treated cases
to the Standard Model predictions represented by an orange dot.

In this way, we present on the upper-left panel of the figure the variations of the
$h\to \gamma \, \gamma$ and $h \to Z\, \gamma$ partial widths for non-vanishing $CP$-conserving
and $CP$-violating parameters $\bar c_{\sss \gamma}$ and $\tilde c_{\sss \gamma}$. We observe that
those two channels do not offer a clear way to distinguish contributions arising from the two
associated ${\cal O}_{\sss \gamma}$ and $\tilde{\cal O}_{\sss \gamma}$ operators. The situation
is however drastically different when we focus, in the
upper-right panel of the figure, on correlations possibly observable
in the $h\to \gamma \, \gamma$ and $h \to Z\, Z$
channels and that are induced by variations of the same $\bar c_{\sss \gamma}$ and $\tilde c_{\sss \gamma}$
parameters. In this case, if one assumes precise measurements
pointing towards physics beyond the Standard Model in one or the other Higgs
decay modes (possibly in collisions
at future LHC center-of-mass energies or even at future colliders), it can be observed
that these measurements could help to get an handle on
the structure of the operators responsible for the assumed observations.
More involved cases are treated in the three other panels of the figure, where we confront predictions
for several partial widths of the Higgs boson into different gauge-boson pairs for various choices of
non-zero Wilson coefficients that are allowed to vary in the $[-1,1]$ range.
For most of the cases, it is found that this method allows
for distinguishing among the different operators that could model a specific new physics effect.

 \subsection{Higgs boson production in association with a vector boson}\label{sec:Htobb}

We now turn to the study of the production of a Higgs boson in association with a weak gauge boson
\be
  p p \to V^* \to h \, V \ ,
\label{eq:htovh}\ee
with $V$ being either a $W$-boson or a $Z$-boson. The existing searches are classified according
to the lepton multiplicity of the final state, \ie, in dileptonic, singly leptonic and zero-lepton
channels. This offers the possibility to make use of the properties of the final state leptons possibly
coming accompanied by missing transverse energy to reject the Standard Model background without
affecting too much the Higgs signal as the latter could be searched for in its dominant decay mode
to two jets originating from the fragmentation of $b$-quarks.

In this case, the final state kinematics differs from the one expected from gluon fusion which was
discussed in the previous subsection~\cite{Djouadi:2013yb,Englert:2013opa}.
In gluon fusion, the Higgs is usually produced with
a very little boost as the partonic center-of-mass energy is close to the Higgs-boson mass.
Contrary, a two-body system comprised of a Higgs boson and a $W$- or $Z$-boson is issued from an off-shell
vector boson that can be seen as radiating a Higgs particle. This off-shellness allows for greater
partonic center-of-mass energies, which subsequently increases the sensitivity to non-standard
Lorentz structures in the interactions of the Higgs field as modeled by the effective operators of
Section~\ref{sec:HELgauge}. We exploit these features in the rest of this
subsection.

\subsubsection{New physics effects in total cross sections for associated Higgs and gauge boson production}
\begin{figure}[t]
  \centering
  \includegraphics[width=.49\columnwidth]{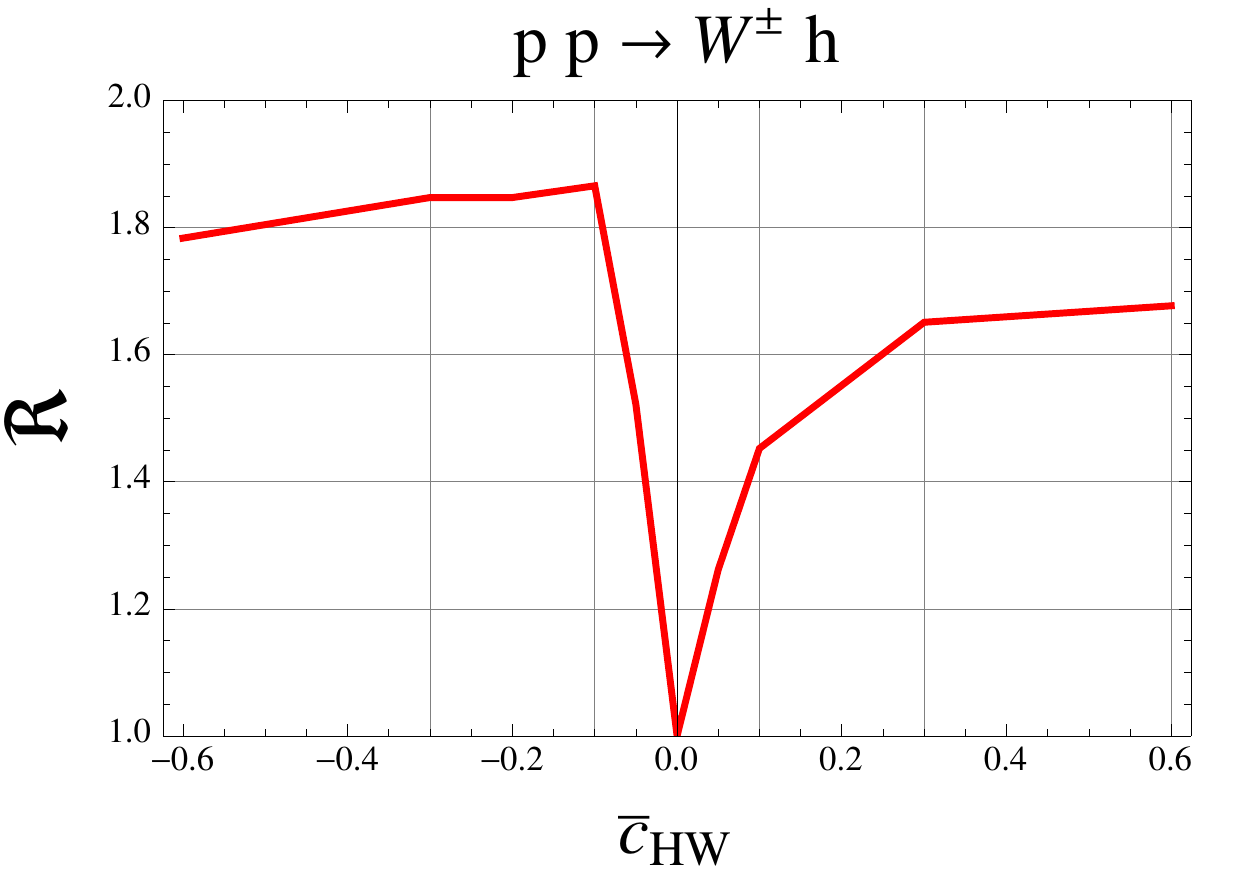}
\caption{
{\it Double ratio ${\cal R}$ of total cross sections at $\sqrt{S} = 8$~TeV and 14~TeV,
  as defined in Eq.~\eqref{eq:doubleR}, given as a function of the value of the $\bar{c}_{\sss HW}$
  parameter for the process $p p \to W^\pm h\to \ell\nu b \bar b$ at the LHC.
  No selection on the final state lepton and missing energy has been accounted for.}}
\label{fig:doubleR}
\end{figure}

Total rates in the associated production of a Higgs boson with a gauge boson~\cite{Ellis:2012xd,%
Ellis:2013ywa,Godbole:2013saa,Isidori:2013cga}
are known to be a powerful handle to obtain
information on beyond the Standard Model effects modeled by dimension-six operators.
In particular, a class of variables consisting of double ratios of total cross sections
has been recently identified as one of the key players for this task~\cite{Mangano:2012mh,Ellis:2013ywa}.
More into details, such quantities are defined
as ratios of total cross sections at different center-of-mass energies,
normalized to the corresponding
Standard Model values. Focusing on present and future LHC collision
center-of-mass energies ($\sqrt{S}~=~8$~TeV and 14~TeV), we investigate in this work
the variable
\be
  {\cal R} \; \equiv \;
    \left(\frac{\sigma (\sqrt{S}= 14 \textrm{ TeV })}{\sigma (\sqrt{S}= 8 \textrm{ TeV })} \right)_{\bar{c}_{\sss i}} /
    \left(\frac{\sigma (\sqrt{S}= 14 \textrm{ TeV })}{\sigma (\sqrt{S}= 8 \textrm{ TeV })} \right)_{SM} 
\label{eq:doubleR} \ee
where the subscript $\bar{c}_{\sss i}$ indicates a computation of the cross section
after including the
effects of an higher-order operator associated with the Wilson coefficient $\bar{c}_{\sss i}$.
It is also believed that this type of variables could be useful in the case
statistically relevant and separated information for collisions at center-of-mass energies
of 13.5~TeV and 14~TeV would be available~\cite{Boudjema:2013qla}.

We illustrate the use of the variable introduced in Eq.~\eqref{eq:doubleR} in
Figure~\ref{fig:doubleR} by investigating the process
\be
  p p \to W^\pm h \to (\ell \nu) \, (b \bar b)\ .
\ee
We show the dependence of ${\cal R}$ on the coefficient of
$\bar{c}_{\sss HW}$ which turns out to be quite steep when $\bar{c}_{\sss HW}$ is of order ${\cal O}(0.1)$
or smaller, and smoother for larger (absolute) values of this Wilson coefficient.
The results however largely depend on the selection requirements
(on the final state lepton and missing transverse energy)
of the corresponding analysis that could further accentuate the effect of the effective operator.

\subsubsection{Invariant mass of a two-body system constituted of a Higgs boson and a gauge boson}\label{sec:mVh}
\begin{figure}[t]
  \centering
  \includegraphics[width=.49\columnwidth]{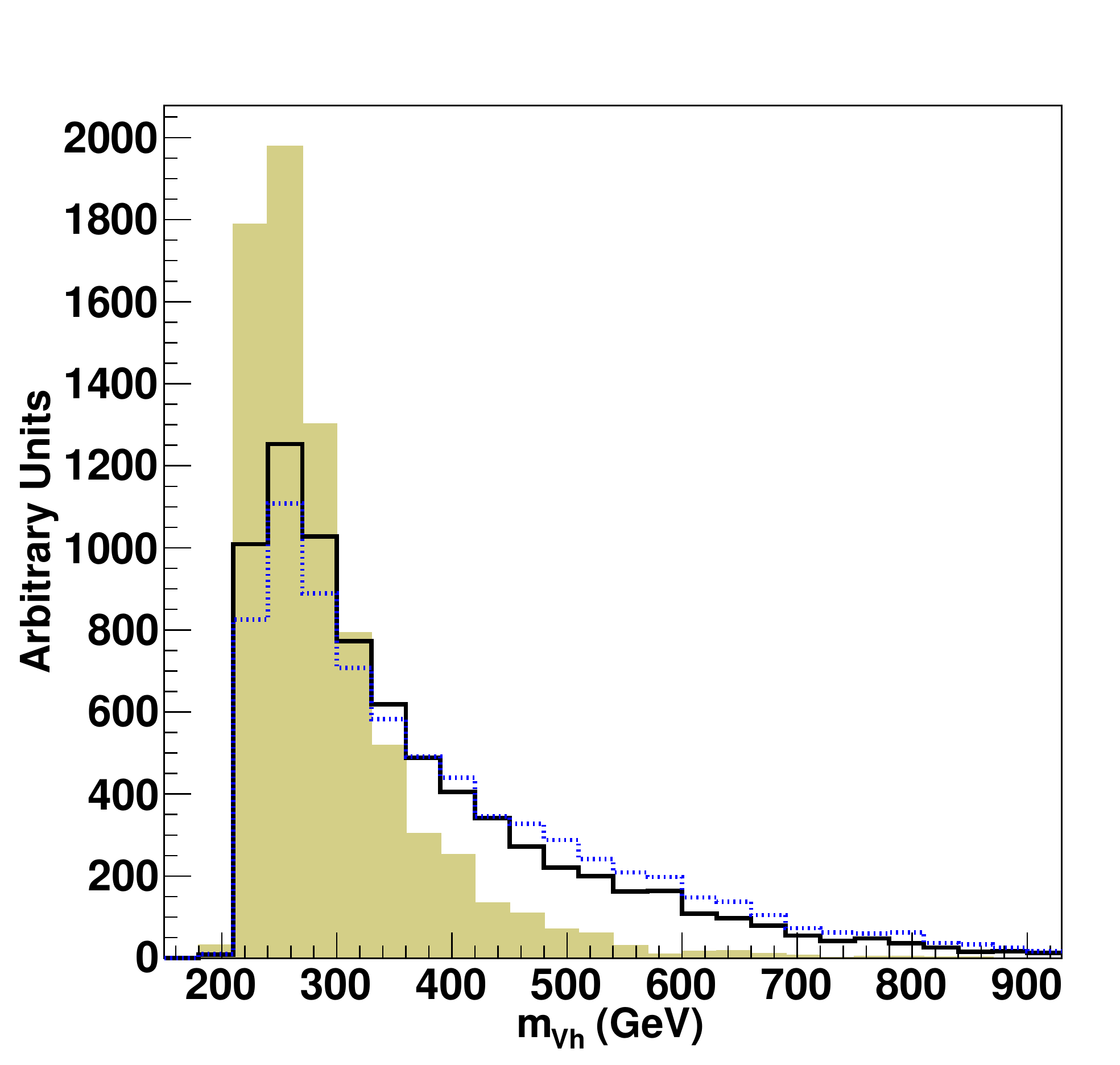}
\caption{
 {\it Invariant-mass $m_{Vh}$ distribution of a two-body system
  comprised of a Higgs boson and a gauge boson for LHC collisions at a center-of-mass energy of 14~TeV.
  We show results for the Standard Model (red-solid histogram) to which we superimpose predictions computed when
  $\bar{c}_{HW}=0.1$ (blue-dotted line) and $\bar{c}_{W}=0.1$ (black-solid line) couplings are allowed.}}
\label{fig:mVH}
\end{figure}

The kinematical properties of the system formed by the massive vector boson
$V$ and the Higgs boson $h$ may be modified by the presence in the Lagrangian
of non-standard operators such as those introduced
in Section~\ref{sec:HELgauge}. In this context, one interesting
observable consists of the invariant-mass distribution
of the $Vh$-system~\cite{Ellis:2012xd}, as illustrated
in Figure~\ref{fig:mVH} for proton-proton collisions at a center-of-mass
energy of 14~TeV. We present in this figure invariant-mass $m_{Vh}$ spectra
computed at the parton-level, \ie, without accounting for gauge-boson and Higgs-boson
decays, and compare the Standard Model predictions (red-solid histogram) to
results including first new physics effects induced by
a non-zero $\bar c_{\sss W}$~=~0.1 parameter
(black-solid line) and second by a non-zero $\bar c_{\sss HW} = 0.1$ parameter (blue-dotted
line). While the Standard Model expectation steeply falls for invariant mass larger
than 500~GeV$-$600~GeV, beyond the Standard Model results exhibit a tail
extending up to much larger $m_{Vh}$ values around the TeV scale.
New operators indeed contribute to this process with different kinematics, favoring configurations with
larger four-momentum. This example therefore
demonstrates the powerful usage of such an observable for unraveling
new physics in the Higgs sector.

\subsection{Di-Higgs production in vector boson fusion}\label{dihiggs}
\begin{figure}[t]
  \centering
  \includegraphics[width=.49\columnwidth]{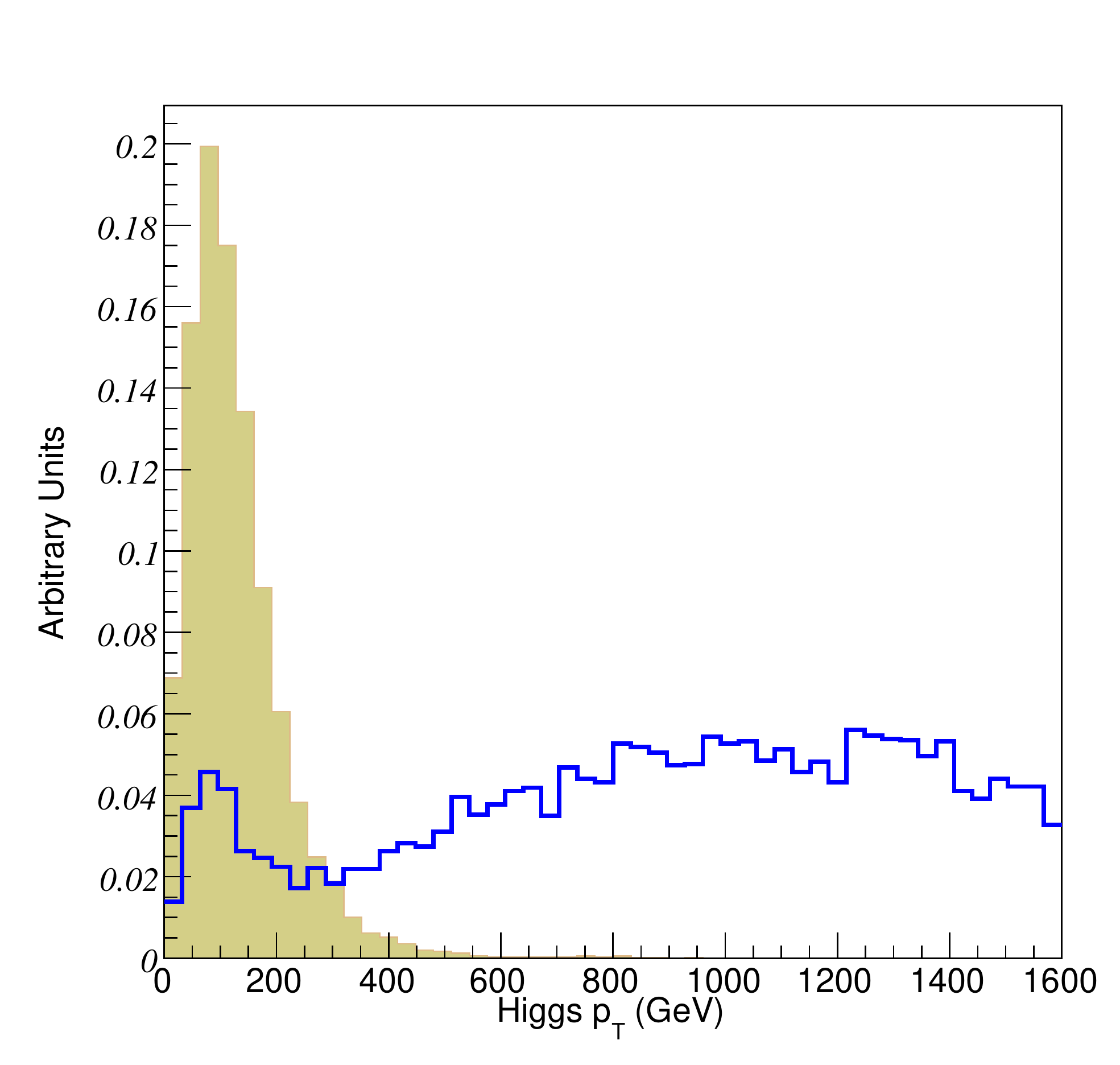}
\caption{
 {\it Transverse-momentum spectrum of both Higgs bosons of a $hh$ pair produced by vector boson fusion,
  the results including hence two entries for each event.
  The green-solid histogram depicts the SM predictions, while the blue-solid line
  corresponds to the addition of new physics effects modeled by $\bar c_{\sss W}=0.05$.}}
\label{fig:Hpt}
\end{figure}

Recently, the interest for di-Higgs production at the LHC, running at a center-of-mass energy of 14~TeV,
has importantly increased. This process indeed allows to get a first grip on the triple-Higgs interaction
strength. In the following, we show as an example di-Higgs production
in the vector boson fusion mode,
\be
  p \, p \to h \, h \, j \, j \ ,
\ee
where both jets $j$ are forward jets. Several of the effective operators of Section~\ref{sec:HELgauge}
can affect such a process and we find that the distribution of
the Higgs transverse momentum $p_T$ provides
information allowing to probe such effects. This is illustrated
on Figure~\ref{fig:Hpt} where we represent the $p_T$ spectrum of both Higgs bosons, including hence
one entry for each Higgs boson in the histograms. We compare the Standard Model expectation (green-solid
histogram) to new physics results arising from $\bar c_{\sss W}=0.05$ (blue-solid line).

We demonstrate in this way how the operator ${\cal O}_{\sss W}$ (and the associated Wilson
coefficient $\bar c_{\sss W}$) favors final state configurations with boosted Higgs bosons.
This behavior is even more pronounced than the one observed in Section~\ref{sec:mVh} when
the Higgs boson is produced in association with a massive gauge boson. As a consequence, the operator
${\cal O}_{\sss W}$ is likely to be investigated via new techniques dedicated to Higgs searches in
boosted topologies, as presented for example in Refs.~\cite{Dolan:2012rv,Dolan:2012ac,Gouzevitch:2013qca,%
Dolan:2013rja}.

\subsection{Associated production of a Higgs and gauge boson from contact interactions with fermions} \label{sec:HZ}
As a last example of the strength of our implementation, we investigate,
in this subsection, possible effects originating
from effective operators
involving one single Higgs field, one single gauge boson and a fermion-antifermion pair.
More especially, we focus on the
${\cal O}_{\sss HQ}$ operator of Eq.~\eqref{eq:lf1} which
contributes
to the associated production of a Higgs boson $h$ together
with a massive vector boson $V$ through a contact interaction,
\be
  p\, p \to h\,  V \ .
\ee
The kinematical properties due to new physics contributions
are thus expected to be largely different from the Standard Model ones where the final state is produced
via Higgs radiation from an off-shell gauge boson, as shown in Eq.~\eqref{eq:htovh}.
Therefore, observables such as the transverse momentum of the Higgs boson or the
invariant mass of the $Vh$ system can be foreseen to play key roles in the detection of
effects due to a non-vanishing $\bar c_{\sss HQ}$ parameter. We have however
found that this operator induces more striking modifications of
the total production cross section, turning thus out to be more promising for
searches for this type of physics beyond the Standard Model.

Considering Higgs boson production in association with a $Z$-boson, we restrict our analysis
to a dileptonic $Z$-decay into an electron or muon pair, together with a Higgs boson decay
into a pair of $b$-quarks,
\be
  p\,  p \to h \,  Z \to (b \, \bar b)\,  (\ell^+\,  \ell^-) \ .
\ee
Simulating LHC collisions at a center-of-mass energy at 14~TeV, we impose
that the four produced fermions have a transverse momentum larger than 25~GeV,
a pseudorapidity satisfying $|\eta|> 2.5$ and that the angular separation $\Delta R$ between
any two objects is greater than 0.4. Normalizing the associated total production rate
to the Standard Model one $\sigma_{SM}$, one fits the effects of the ${\cal O}_{\sss HQ}$
operator as
\be
  \kappa_{\bar c_{\sss HQ}} = \frac{\sigma_{\bar c_{\sss HQ}}}{\sigma_{SM}} =
    1.00 - 2.00  \, \bar c_{\sss HQ} + 863 \bar c_{\sss HQ}^2\ ,
\ee
where $\sigma_{\bar c_{\sss HQ}}$ stands for the cross section as computed when including,
in addition to the SM contributions, diagrams involving the ${\cal O}_{\sss HQ}$ effective
operator. This proves that any LHC measurement of this $\kappa_{\bar c_{\sss HQ}}$ quantity
that is accurate at the percent level (achievable
with the high-luminosity run of the LHC~\cite{Butterworth:2008tr}) would surpass the
current bounds derived from LEP-I data at the Z-pole, the latter constraining
the value of $\bar c_{HQ}$ to be of order ${\cal O}(10^{-3})$~\cite{Pomarol:2013zra}.

\section{Conclusions}\label{sec:concl}

The Higgs discovery, together with the characterization of its properties and quantum numbers,
sets the ground for the approach taken in this paper. With the absence of any evidence for new physics,
we have followed the path of effective field theories to describe
beyond the Standard Model effects in the Higgs sector. We have first formulated in extensive
details the Higgs Effective Lagrangian that we have employed, limiting ourselves
to dimension-six operators involving the Higgs and/or gauge bosons and adopting
the basis and conventions of Refs.~\cite{Giudice:2007fh,Contino:2013kra}.
Next, we have implemented this Lagrangian in the framework of {\sc FeynRules}, a {\sc Mathematica}
package interfaced to several sophisticated Monte Carlo tools.
Our implementation allows to study, in particular, various differential distributions related
to processes involving one or several Higgs bosons in the context of the LHC collisions.
In the high-energy run of the LHC at a center-of-mass energy of 14~TeV, more statistics will be collected,
resulting in a better understanding of the Higgs properties. In these perspectives, fits involving the Higgs signal
strengths will rapidly fall short to exploit all data, while angular, invariant-mass as well as
many other differential distributions will be able to directly probe possible
non-standard Lorentz structures of the Higgs interactions. We believe
that our implementation will hence allow to unveil physics beyond the Standard Model
if related to the Higgs sector.

We have illustrated the strengths of our machinery by making use of the UFO
interface of {\sc FeynRules} to pass the full set of interaction vertices included in
our Higgs Effective Lagrangian to the event generator {\sc MadGraph}~5, without
any restriction on the Lorentz structures or on the number of external legs allowed in the vertices.
For the sake of the example, we have
considered Higgs boson production via gluon fusion and investigated its decay modes
to massive vector bosons, Higgs boson associated production with a weak
boson, di-Higgs production as well as the effects of contact interactions among
a fermion-antifermion pair, a Higgs boson and a gauge boson. In all those processes,
we have studied either total rates, or several differential distributions, or both, and shown
how this has allowed us to investigate how dimension-six operators
could manifest themselves in LHC processes. Finally, we have also briefly addressed
the way to disentangle the source of a specific
effect among the set of effective operators possibly giving rise to it.

\acknowledgments
The authors are grateful to Fabio Maltoni and Kentarou Mawatari for
enlightening discussions during all the phases of this project.
The work of VS has been supported by the Science Technology and Facilities Council (STFC)
under grant number ST/J000477/1, while the work of AA and BF has received partial support
from the Theorie-LHC France initiative of the CNRS/IN2P3.

\bibliographystyle{JHEP}
\bibliography{biblio}

\end{document}